\documentclass[12pt,prd]{revtex4-1}
\usepackage{CJK}
\usepackage[T1]{fontenc}
\usepackage[left=2.5cm,top=3.0cm,right=2.5cm,bottom=3.0cm]{geometry} 
\usepackage[utf8]{inputenc} 

\usepackage{array}
\usepackage{textcomp}
\usepackage{multirow}
\usepackage{amsmath}
\usepackage{graphicx}
\usepackage{hhline}
\usepackage{breqn}
\usepackage{comment}
\usepackage{adjustbox}
\usepackage{support-caption}
\usepackage{subcaption}
\usepackage{appendix}
\usepackage{slashed}

\usepackage{amsmath}
\usepackage{amsfonts}
\usepackage{amssymb}
\usepackage{cancel} 
\usepackage{youngtab}
\usepackage{xfrac}

\usepackage{pstricks}
\usepackage{color}

\usepackage{graphicx}
\usepackage{feynmp} 
\usepackage{float} 

\newcommand{\be}{\begin{equation}}
\newcommand{\ee}{\end{equation}}
\newcommand{\bea}{\begin{eqnarray}}
\newcommand{\eea}{\end{eqnarray}}

\usepackage{booktabs}

\begin{document}

\title{$\tilde{\nu}$ contributions to electron and muon EDM in an Inverse Seesaw Mechanism.}
\author{J.S. Alvarado} \thanks{jsalvaradog@unal.edu.co}
\author{R. Martinez. } \thanks{remartinezm@unal.edu.co}

\affiliation{Departamento de Física$,$ Universidad Nacional de Colombia\\
 Ciudad Universitaria$,$ K. 45 No. 26-85$,$ Bogotá D.C.$,$ Colombia}
\date{\today}

\begin{abstract}
A non-universal anomaly free $U(1)_{X}$ extension to the Minimal Supersymmetric Standard Model, consisting of four scalar doublets, four scalar singlets and additional quark and lepton singlets including right-handed and Majorana neutrinos, is used to determine the contributions to the electron and muon Electric Dipole Moment. The additional CP violation sources come from the lepton sector, where neutrino masses are explained by an Inverse Seesaw Mechanism and the CP violating phase of the PMNS matrix generates complex interactions that involves exotic neutrino and sneutrino mass eigenstates. Such contributions are studied at one and two-loop level by considering the associated Barr-Zee diagrams and their supersymmetric counterpart. At one-loop level, it is found that the Electric Dipole Moment fixes a relationship between chargino and sneutrino masses depending on which particles have a mass bellow $10^{6}$ GeV. At two loop level, contributions are comparable to the one-loop contributions but the integrals diverge in some cases, yielding additional restrictions such as no degenerate sneutrino masses and they should be heavier than chargino masses.
    
\end{abstract}

\maketitle

\section{Introduction}
Despite a CP-violation source is already known in the quark sector of the Standard Model (SM), it is not able to explain the cosmic baryon asymmetry of the universe so additional CP violation sources are expected and searched as well. Particularly, a non-zero EDM of any elementary particle would undoubtedly imply new sources of CP violation beyond the Standard Model of particle physics. In the SM, the CKM matrix predicts at four-loop  $|d_{e}|\sim 10^{-44} \; e \; cm$ \cite{barr-marciano} and a CP-odd electron-nucleon interaction whose prediction is $|d_{e}|\sim 10^{-38} \; e\; cm$ \cite{CPoddnucleon}. Nevertheless, it has been recently proved that the hadron level long distance effect generates a large EDM when considering vector meson loops, providing a value of $5.8\times 10^{-40} e\; cm$ for the electron and $1.4\times 10^{-38} e\; cm$ for the muon with a theoretical uncertainty around $70\%$ \cite{yamanaka1}\cite{yamanaka2}. However, additional CP-violation sources such as the strong $\bar{\theta}$ phase has a prediction of $|d_{e}|\leq 10^{-37} \; e \; cm$ \cite{strongtheta} and CP violation in the lepton sector has a null contribution in the SM \cite{electronEDMzero} and small predictions of $|d_{e}| \leq 10^{-43}$ when Majorana neutrinos are considered \cite{MajoranacontributiontoEDM} . 

Currently, the electron EDM upper bound is set by the ThO experiment of ACME collaboration \cite{ACMEexp} that reports  $d_{e}  \leq 1.1 \times 10^{-29} \;e\;cm$ at 90\% confidence level (C.L.) in agreement with HfF$^+$ at JILA studies \cite{JILA}. However, an important sensitivity improvement is expected from the EDM$^3$ experiment in the near future \cite{EDM3}. Additionally, muon EDM upper bound is given by $ d_{\mu}< 1.5 \times 10^{-19}$ e cm at  $90\% $  C.L. \citep{muonEDM} by the Muon $g-2$ collaboration although there is a proposed experiment with a frozen-spin technique at PSI that could perform muon EDM searches with a sensitivity of $\sigma(d_{\mu})<6\times 10^{-23} e\;cm$ \cite{PSI}.

Moreover, neutrino masses are another promising new physics problem  since neutrino oscillation was confirmed \cite{nuoscillations}, leading to different mass generation mechanisms such as the seesaw models \cite{seesaw}\cite{inverseseesaw}. Such models consider additional heavy particles such as right-handed Majorana neutrinos or several additional sterile neutrinos which in general may have complex couplings to explain the CP phase of the PMNS matrix. Since such particles must have considerably heavy masses, contrary to SM neutrinos, their contribution to EDMs would be no longer negligible. 

The EDM has been considered a long time ago taking into account the Barr-Zee diagrams at two-loop \cite{barr-zee} in CP violating Higgs sector \cite{CPviolatingHiggsSector}, 3-gluon operators \cite{3gluon} or the Two Higgs Doublet Model among others \cite{2HDMEDM}. In the case of the muon, its anomalous magnetic dipole moment raises the question about the implication on its EDM due to possible beyond the Standard Model effects as pointed out in \citep{muonMDMEDM}. Furthermore, the supersymmetric scenario has been widely studied as well, first focused on the neutron EDM \cite{SUSYneutron} then on lepton EDM by considering stop particles \cite{stop}, CP violation coming from soft SUSY breaking and theories beyond the Minimal Supersymetric Standard Model (MSSM) \cite{BMSSM} such as the BLMSSM \cite{BLMSSM} and the R-parity violating MSSM \cite{yamanaka3}.

The present work considers a non-universal $U(1)_{X}$ extension to the MSSM, consisting of four scalar doublets and four scalar singlets among other fermion singlets, which provides an explanation for fermion mass hierarchy \cite{model}, it is compatible with the PMNS matrix elements \cite{modelPMNS} and can explain the muon $g-2$ anomaly \cite{modelg-2}, in an scenario where neutrino masses are explained by an Inverse Seesaw Mechanism. However, just like exotic neutrinos might have important contributions, the supersymmetric scenario implies that sneutrino contributions might be important as well.

\section{The $U(1)_{X}$ extension}
 
 The proposed model considers an additional $U(1)_{X}\times \mathcal{Z}_{2}$ global symmetry to the MSSM with non-universal $X$ charge and parity assignation that generates a zero-texture mass matrices compatible with fermion masses. A total of four scalar doublets and four scalar singlets make up the scalar sector, shown in table \ref{modelbosons}, whose Vacuum Expectation Values (VEV) provide a mechanism for understanding fermion mass hierarchy under the spontaneous symmetry breaking chain: 
 
 \begin{equation}
    \mathrm{SU(3)}_{C}\otimes
    \mathrm{SU(2)}_{L}\otimes 
    \mathrm{U(1)}_{Y} \otimes 
    \mathrm{U(1)}_{X} \overset{\chi}{\longrightarrow}
    \mathrm{SU(3)}_{C}\otimes
    \mathrm{SU(2)}_{L}\otimes 
    \mathrm{U(1)}_{Y} \overset{\Phi}{\longrightarrow}
    \mathrm{SU(3)}_{C}\otimes
    \mathrm{U(1)}_{Q} \nonumber
\end{equation}

\noindent
 being the scalar singlets $\chi$ and $\chi'$ responsible of the $U(1)_{X}$ symmetry breaking and the scalars $\sigma$ and $\sigma'$ make the lightest fermions massive at one-loop level. The fermion sector, shown in table \ref{modelfermions}, comprise an additional up-quark singlet ($\mathcal{T}$), two down-like quark singlets ($\mathcal{J}^{a}$, $a=1,2$), two charged lepton singlets ($E$, $\mathcal{E}$), three right-handed neutrinos ($\nu_{L}^{C}$) and three heavy Majorana neutrinos ($N_{R}$). All exotic particles have an expected big mass which is justified by either $\chi$ or $\chi^{\prime}$. In particular, $Z'$ gauge boson mass can be approximated to $M_{Z'}\approx \sfrac{g_{X}\sqrt{v_{\chi}^{2}+v_{\chi}^{\prime2}}}{3}$ so it is reasonable to think of $v_{\chi}$ and $v_{\chi}^{\prime}$, at least, at the TeV scale. Moreover, neutrino masses are explained by an Inverse Seesaw Mechanism resulting in the three active SM neutrinos and six heavy Majorana neutrino eigenstates \cite{model}.

Nevertheless, despite the new $U(1)_{X}$ symmetry might induce undesirable chiral anomalies, the charge assignation does vanish the anomaly equations shown in Eqs. (\ref{an1})-(\ref{an6}), leaving the model anomaly free

\begin{eqnarray}
\left[\mathrm{\mathrm{SU}(3)}_{C} \right]^{2} \mathrm{\mathrm{U}(1)}_{X} \rightarrow & A_{C} &= \sum_{Q}X_{Q_{L}} + \sum_{Q}X_{Q_{L}^{c}},	\label{an1}	\\
\left[\mathrm{\mathrm{SU}(2)}_{L} \right]^{2} \mathrm{\mathrm{U}(1)}_{X} \rightarrow & A_{L}  &= \sum_{\ell}X_{\ell_{L}} + 3\sum_{Q}X_{Q_{L}}	,	\\
\left[\mathrm{\mathrm{U}(1)}_{Y} \right]^{2}   \mathrm{\mathrm{U}(1)}_{X} \rightarrow & A_{Y^{2}}&=
	\sum_{\ell, Q}\left[Y_{\ell_{L}}^{2}X_{\ell_{L}}+3Y_{Q_{L}}^{2}X_{Q_{L}} \right]	
	+ \sum_{\ell,Q}\left[Y_{\ell_{L}^{c}}^{2}X_{L_{L}^{c}}+3Y_{Q_{L}^{c}}^{2}X_{Q_{L}^{c}} \right],		\\
\mathrm{\mathrm{U}(1)}_{Y}   \left[\mathrm{\mathrm{U}(1)}_{X} \right]^{2} \rightarrow & A_{Y}&=
	\sum_{\ell, Q}\left[Y_{\ell_{L}}X_{\ell_{L}}^{2}+3Y_{Q_{L}}X_{Q_{L}}^{2} \right]	
	+ \sum_{\ell, Q}\left[Y_{\ell_{L}^{c}}X_{\ell_{L}^{c}}^{2}+3Y_{Q_{L}^{c}}X_{Q_{L}^{c}}^{2} \right],		\\
 \left[\mathrm{\mathrm{U}(1)}_{X} \right]^{3} \rightarrow & A_{X}&=
	\sum_{\ell, Q}\left[X_{\ell_{L}}^{3}+3X_{Q_{L}}^{3} \right]	
	+ \sum_{\ell, Q}\left[X_{\ell_{L}^{c}}^{3}+3X_{Q_{L}^{c}}^{3} \right] 	,	\\	
\left[\mathrm{Grav} \right]^{2}   \mathrm{\mathrm{U}(1)}_{X} \rightarrow & A_{\mathrm{G}}&=
	\sum_{\ell, Q}\left[X_{\ell_{L}}+3X_{Q_{L}} \right]
	+ \sum_{\ell, Q}\left[X_{\ell_{L}^{c}}+3X_{Q_{L}^{c}} \right] \label{an6}.
\end{eqnarray}

On the other hand, the additional quantum number does not affect the definition of electric charge so its definition is given by the Gell-Mann-Nishijima relationship , $Q=\mathcal{I}_{3}+\frac{1}{2}Y$. Besides, it is worth to mention that right handed fields are represented by left-conjugate ones ($\bar{\psi_{R}}\rightarrow \psi_{L}^{c}$) making right handed particles in the model to present the opposite electric charge.

\begin{table}[H]
\caption{Scalar content of the model, hypercharge $Y$, non-universal $X$ quantum number and $\mathbb{Z}_{2}$ written in the form $X^{\mathbb{Z}_{2}}$.}
\label{modelbosons}
\centering
\begin{tabular}{lll cll}\hline\hline 
\multirow{3}{*}{
\begin{tabular}{l}
    Higgs    \\
    Scalar  \\
    Doublets
\end{tabular}
}
&\multicolumn{2}{l}{}&
\multirow{3}{*}{
\begin{tabular}{l}
    Higgs    \\
    Scalar  \\
    Singlets
\end{tabular}
} 
&\multicolumn{2}{l}{}\\ 
 &&&
 && \\ 
 &$X^{\pm}$&$Y$&
 &$X^{\pm}$&$Y$
\\ \hline\hline 
$\small{\hat{\Phi}_{1}=\begin{pmatrix}\hat{\phi}_{1}^{+}\\\frac{\hat{h}_{1}+v_{1}+i\hat{\eta}_{1}}{\sqrt{2}}\end{pmatrix}}$&$\sfrac{+2}{3}^{+}$&$+1$&
$\hat{\chi}=\frac{\hat{\xi}_{\chi}+v_{\chi}+i\hat{\zeta}_{\chi}}{\sqrt{2}}$	&	$\sfrac{-1}{3}^{+}$	&	$0$	\\
$\small{\hat{\Phi}_{2}=\begin{pmatrix}\hat{\phi}_{2}^{+}\\\frac{\hat{h}_{2}+v_{2}+i\hat{\eta}_{2}}{\sqrt{2}}\end{pmatrix}}$&$\sfrac{+1}{3}^{-}$&$+1$& $\sigma=\frac{\hat{\xi}_{\sigma}+i\hat{\zeta}_{\sigma}}{\sqrt{2}} $ &$ \sfrac{-1}{3}^{-} $ & $ 0 $		\\
$\small{\hat{\Phi}^\prime_{1}=\begin{pmatrix}\frac{\hat{h}_{1}'+v_{1}'+i\hat{\eta}_{1}'}{\sqrt{2}}\\\hat{\phi}_{1}^{-\prime}\end{pmatrix}}$&$\sfrac{-2}{3}^{+}$&$-1$&
$\hat{\chi}'=\frac{\hat{\xi}'_{\chi}+v_{\chi}'+i\hat{\zeta}'_{\chi}}{\sqrt{2}}$	&	$\sfrac{+1}{3}^{+}$ &	0\\
$\small{\hat{\Phi}^\prime_{2}=\begin{pmatrix}\frac{\hat{h}_{2}'+v_{2}'+i\hat{\eta}_{2}'}{\sqrt{2}}\\\hat{\phi}_{2}^{-\prime}\end{pmatrix}}$&$\sfrac{-1}{3}^{-}$&$-1$& $\sigma^{\prime} = \frac{\hat{\xi}_{\sigma}^{\prime}+i\hat{\zeta}_{\sigma}^{\prime}}{\sqrt{2}}$ & $\sfrac{+1}{3}^{-}$ &	0	\\\hline\hline
\end{tabular}
\end{table}

\begin{table}
\caption{Fermion content of the non-universal extension, hypercharge $Y$, $X$ quantum number and parity $\mathbb{Z}_{2}$ written in the form $X^{\mathbb{Z}_{2}}$.}
\label{modelfermions}
\centering
\begin{tabular}{lll lll}\hline\hline 
 Left-Handed Fermions &$X^{\pm}$&&
 Right-Handed Fermions &$X^{\pm}$&
\\ \hline\hline 
\multicolumn{6}{c}{SM Quarks }\\ 
\multicolumn{6}{c}{$Y_{q_{L}}=\sfrac{+1}{3}$, $Y_{u_{L}^{c}}=\sfrac{-4}{3}$, $Y_{d_{L}^{c}}=\sfrac{+2}{3}$}\\ \hline\hline
\begin{tabular}{c}	
	$ \hat{q} ^{1}_{L}=\begin{pmatrix}\hat{u}^{1}	\\ \hat{d}^{1} \end{pmatrix}_{L}$  \\
	$  \hat{q} ^{2}_{L}=\begin{pmatrix}\hat{u}^{2}	\\ \hat{d}^{2} \end{pmatrix}_{L}$  \\
	$  \hat{q} ^{3}_{L}=\begin{pmatrix}\hat{u}^{3}	\\ \hat{d}^{3} \end{pmatrix}_{L}$ 
\end{tabular} &
\begin{tabular}{c}
		$\sfrac{+1}{3}^{+}$	\\
	\\	$0^{-}$	\\
	\\	$0^{+}$	\\
\end{tabular}   &
\begin{tabular}{c}
			\\
	\\		\\
	\\		\\
\end{tabular}   &
\begin{tabular}{c}
	$ \begin{matrix}\hat{u}^{1\;     c }_{L}	\\ \hat{u}^{2\; c}_{L} \end{matrix}$  \\
	$ \begin{matrix}\hat{u}^{3\; c}_{L}	\\ \hat{d}^{1\; c }_{L} \end{matrix}$  \\
	$ \begin{matrix}\hat{d}^{2\; c }_{L}	\\ \hat{d}^{3\; c }_{L} \end{matrix}$ 
\end{tabular} &
\begin{tabular}{c}
	$ \begin{matrix} \sfrac{-2}{3}^{+}	\\ \sfrac{-2}{3}^{-} \end{matrix}$  \\
	$ \begin{matrix} \sfrac{-2}{3}^{+}	\\ \sfrac{+1}{3}^{-} \end{matrix}$  \\
	$ \begin{matrix} \sfrac{+1}{3}^{-}	\\ \sfrac{+1}{3}^{-} \end{matrix}$ 
\end{tabular} &
\begin{tabular}{c}
	\\
	 \\
\end{tabular}
\\ \hline\hline 

\multicolumn{6}{c}{SM Leptons}\\ 
\multicolumn{6}{c}{$Y_{\ell_{L}}=-1$, $Y_{e_{L}^{c}}=+2$, $Y_{\nu_{L}^{c}}=0$}\\ \hline\hline	
\begin{tabular}{c}	
	$\hat{\ell}^{e}_{L}=\begin{pmatrix}\hat{\nu}^{e}\\ \hat{e} \end{pmatrix}_{L}$  \\
	$\hat{\ell}^{\mu}_{L}=\begin{pmatrix}\hat{\nu}^{\mu}\\ \hat{\mu} \end{pmatrix}_{L}$  \\
	$\hat{\ell}^{\tau}_{L}=\begin{pmatrix}\hat{\nu}^{\tau}\\ \hat{\tau} \end{pmatrix}_{L}$ 
\end{tabular} &
\begin{tabular}{c}
		$0^{+}$	\\
	\\	$0^{+}$	\\
	\\	$-1^{+}$	\\
\end{tabular} &
\begin{tabular}{c}
		\\
	\\		\\
	\\		\\
\end{tabular}   &
\begin{tabular}{c}
	$ \begin{matrix}\hat{\nu}^{e\; c}_{L}	\\ \hat{\nu}^{\mu\; c}_{L} \end{matrix}$  \\
	$ \begin{matrix}\hat{\nu}^{\tau\; c }_{L}	\\  \hat{e}^{e\; c}_{L} \end{matrix}$  \\
	$ \begin{matrix} \hat{e}^{\mu\; c }_{L}	\\  \hat{e}^{\tau\; c}_{L} \end{matrix}$ 
\end{tabular} &
\begin{tabular}{c}
	$ \begin{matrix} \sfrac{-1}{3}^{-}	\\ \sfrac{-1}{3}^{-} \end{matrix}$  \\
	$ \begin{matrix} \sfrac{-1}{3}^{-}	\\ \sfrac{+4}{3}^{-} \end{matrix}$  \\
	$ \begin{matrix} \sfrac{+1}{3}^{-}	\\ \sfrac{+4}{3}^{-} \end{matrix}$ 
\end{tabular} &
\begin{tabular}{c}
	  \\
	 \\
\end{tabular}
\\ \hline\hline 

\multicolumn{6}{c}{Non-SM Quarks: $Y_{\mathcal{T}_{L}}=-Y_{\mathcal{T}_{L}^{c}}=\sfrac{-4}{3}$, $Y_{\mathcal{J}_{L}}=-Y_{\mathcal{J}_{L}^{c}}=\sfrac{+2}{3}$}\\ \hline\hline	
\begin{tabular}{c}	
	$\hat{\mathcal{T}}_{L}$	\\	
	$\mathcal{J}_{L}^{1}$	\\	$\mathcal{J}_{L}^{2}$	
\end{tabular} &
\begin{tabular}{c}
	$\sfrac{+1}{3}^{-} $	\\	
	$ 0^{+} $           	\\	$ 0^{+} $
\end{tabular}   &
\begin{tabular}{c}
		\\		\\	
		\\		
\end{tabular}   &

\begin{tabular}{c}
	$\hat{\mathcal{T}}_{L}^{c}$	\\	
	$\hat{\mathcal{J}}_{L}^{c \ 1}$	\\	$\hat{\mathcal{J}}_{L}^{c \ 2}$	
\end{tabular} &
\begin{tabular}{c}
	$\sfrac{-2}{3}^{-} $	\\	  
	$\sfrac{+1}{3}^{+} $	\\	$\sfrac{+1}{3}^{+} $
\end{tabular} &
\begin{tabular}{c}
		\\		\\	
		\\		
\end{tabular}
\\ \hline\hline

\multicolumn{6}{c}{Non-SM Leptons: $Y_{E_{L}}=-Y_{E_{L}^{c}}=Y_{\mathcal{E}_{L}}=-Y_{\mathcal{E}_{L}^{c}}=-2$}\\ \hline\hline	
\begin{tabular}{c}	
    $\hat{E}_{L}$	\\
    $\hat{\mathcal{E}}_{L}$	\\
\end{tabular} &
\begin{tabular}{c}
	$-1^{+}$	    	\\
	$\sfrac{-2}{3}^{+}$	\\
\end{tabular} &
\begin{tabular}{c}
			\\
			\\
		
\end{tabular}   &
\begin{tabular}{c}	
    $\hat{E}_{L}^{c}$	\\
    $\hat{\mathcal{E}}_{L}^{c}$	\\
\end{tabular} &
\begin{tabular}{c}
	$\sfrac{+2}{3}^{+}$		\\
	$+1^{+}$	            \\
\end{tabular} &
\begin{tabular}{c}
			\\
			\\
	
\end{tabular}
\\	\hline\hline 
\multicolumn{3}{c}{Majorana Fermions: $Y_{\mathcal{N}}=0$} & 
\begin{tabular}{c}	
	$\mathcal{N}_{R}^{1,2,3}$	\\
\end{tabular} &
\begin{tabular}{c}
	$0^{-}$	\\
\end{tabular} &
\begin{tabular}{c}
	\\
\end{tabular}
\\	\hline\hline 
\end{tabular}
\end{table}

Finally, gauge invariance induce the D-term potential shown in Eq. (\ref{VD}) and the superpotential given in Eq. (\ref{VW}) while  SUSY is broken explicitly by the soft breaking potential shown in Eq. (\ref{Vs}). The latter allows the presence of a $125\; GeV$ scalar compatible with the Higgs boson as it can be detailed seen in \cite{model}. Furthermore, it provides the masses of charginos, neutralinos and sparticles as free parameters since their energy scale would be expected at least at the TeV scale, making all superpotential and D-term contributions negligible in comparison. Such potentials read:

\begin{small}
\begin{align}
V_{D}&=\frac{g^{2}}{2}\Big[ |\Phi_{1}^{\dagger}\Phi_{2}|^{2}+|\Phi_{1}^{\prime\dagger}\Phi_{2}'|^2+|\Phi_{1}^{\prime\dagger}\Phi_{1}|^2+|\Phi_{1}^{\prime\dagger}\Phi_{2}|^2+|\Phi_{2}^{\prime\dagger}\Phi_{1}|^2+|\Phi_{2}^{\prime\dagger}\Phi_{2}|^2\nonumber\\
 &-|\Phi_{1}|^{2}|\Phi_{2}|^{2}-|\Phi_{1}^{\prime}|^{2}|\Phi_{2}^{\prime}|^{2} \Big]+\frac{g^{2} + g^{\prime 2}}{8}(\Phi_{1}^{\dagger}\Phi_{1}+\Phi_{2}^{\dagger}\Phi_{2}-\Phi_{1}^{\prime\dagger}\Phi_{1}^{\prime}-\Phi_{2}^{\prime\dagger}\Phi_{2}^{\prime})^{2} \nonumber\\
 &+\frac{g_{X}^{2}}{2}\left[\frac{2}{3}(\Phi_{1}^{\dagger}\Phi_{1}-\Phi_{1}^{\prime\dagger}\Phi_{1}^{\prime})+\frac{1}{3}(\Phi_{2}^{\dagger}\Phi_{2}-\Phi_{2}^{\prime\dagger}\Phi_{2}^{\prime})-\frac{1}{3}(\chi^{*}\chi-\chi^{\prime*}\chi^{\prime}) -\frac{1}{3}(\sigma^{*}\sigma-\sigma^{\prime*}\sigma^{\prime})\right]^{2}, \label{VD} \\
W[\phi]&=-\mu_{1}\hat{\Phi}'_{1}\hat{\Phi}_{1}-\mu_{2}\hat{\Phi}'_{2}\hat{\Phi}_{2} - \mu_{\chi}\hat{\chi} '\hat{\chi} - \mu_{\sigma}\hat{\sigma} '\hat{\sigma} + \lambda_{1}\hat{\Phi}_{1}^{\prime}\hat{\Phi}_{2}\hat{\sigma}^{\prime} + \lambda_{2}\hat{\Phi}_{2}^{\prime}\hat{\Phi}_{1}\sigma, \label{VW} \\
    V_{soft}&=m_{1}^{2}\Phi_{1}^{\dagger}\Phi_{1} + {m}_{1}^{\prime 2}{\Phi}_{ 1}^{\prime \dagger}\Phi'_{1} + m_{2}^{2}\Phi_{2}^{\dagger}\Phi_{2} + {m}_{2}^{\prime 2}\Phi _{2}^{\prime\dagger}\Phi'_{2}+m_{\chi}^{2}\chi^{\dagger}\chi + {m}_{\chi}^{\prime 2}{\chi}^{\prime\dagger}\chi'+m_{\sigma}^{2}\sigma^{\dagger}\sigma \nonumber\\
    & + {m}_{\sigma}^{\prime 2}{\sigma}^{\prime\dagger}\sigma'  -\bigg[\mu_{11}^{2}\epsilon_{ij}({\Phi}_{1}^{\prime i}\Phi_{1}^{j}) -\mu_{22}^{2}\epsilon_{ij}({\Phi}_{2}^{\prime i}\Phi_{2}^{j}) -\mu_{\chi\chi}^{2}(\chi\chi') +\mu_{\sigma\sigma}^{2}(\sigma\sigma') + \tilde{\lambda}_{1}\Phi_{1}^{\prime \dagger}\Phi_{2}\sigma^{\prime}\nonumber\\
    & + \tilde{\lambda}_{2}\Phi_{2}^{\prime \dagger}\Phi_{1}\sigma - \frac{2\sqrt{2}}{9}(k_{1}\Phi_{1}^{\dagger}\Phi_{2}\chi' -k_{2}\Phi_{1}^{\dagger}\Phi_{2}\chi^*+k_{3}\Phi_{1}'{}^{\dagger}\Phi_{2}'\chi -k_{4}\Phi_{1}'{}^{\dagger}\Phi_{2}'\chi'{}^*)+h.c.\bigg] \nonumber \\
    &+ M_{\tilde{B}}\tilde{B}\tilde{B}^{\dagger} + M_{\tilde{B}'}\tilde{B}'\tilde{B}^{\prime \dagger} + M_{\tilde{W}^{\pm}}\tilde{W}^{\pm}\tilde{W}^{\pm \dagger} + M_{\tilde{W}}\tilde{W}_{3}\tilde{W}_{3}^{\dagger} + \sum_{\tilde{f} \in sparticles}m_{k}^{2} \tilde{f}\tilde{f}^{\dagger},\label{Vs}
\end{align}
\end{small}

\noindent
where the terms proportional to $k_{1},k_{2},k_{3}$ and $k_{4}$ softly break the parity symmetry.  Finally, considering the potential due to F-terms  and just taking the contribution to the scalar potential, we obtain:

\begin{eqnarray}\label{VF}
V_{F}&=&\mu_{1}^2 ( \Phi_{1}^\dagger\Phi_{1}+\Phi_{1}^{\prime\dagger}\Phi_{1}^{\prime})+\mu_{2}^2(\Phi_{2}^\dagger\Phi_{2}+\Phi_{2}^{\prime\dagger}\Phi_{2}^{\prime})
+\mu_{\chi}^2(\chi^*\chi+\chi^{\prime*}\chi') + +\mu_{\sigma}^2(\sigma^*\sigma+\sigma^{\prime*}\sigma^{\prime})\nonumber\\
&+& (\lambda_{1}^{2}|\epsilon_{ij}\Phi_{1}^{\prime i}\Phi_{2}^{j}|^{2} + \lambda_{2}^{2}|\epsilon_{ij}\Phi_{2}^{\prime i}\Phi_{1}^{j}|^{2}  + \lambda_{1}^{2}( \Phi_{2}^{\dagger} \Phi_{2} + \Phi_{1}^{\prime \dagger}\Phi_{1}^{\prime}\sigma^{\prime *}\sigma^{\prime} + \lambda_{2}^{2}( \Phi_{1}^{\dagger}\Phi_{1} + \Phi_{2}^{\prime \dagger}\Phi_{2}^{\prime})\sigma^{*}\sigma \nonumber\\
& -&\lambda_{1}\mu_{1}\Phi_{1}^{\dagger}\Phi_{2}\sigma^{\prime} - \lambda_{1}\mu_{2}\Phi_{2}^{\prime \dagger }\Phi_{1}^{\prime}\sigma^{\prime}  -\lambda_{2}\mu_{1}\Phi_{1}^{\prime \dagger}\Phi_{2}^{\prime}\sigma -\lambda_{2}\mu_{2}\Phi_{2}^{\dagger}\Phi_{1}\sigma - \lambda_{1}\mu_{\sigma}\epsilon_{ij}\Phi_{1}^{\prime i}\Phi_{2}^{j} \nonumber\\
& -&\lambda_{2}\mu_{\sigma}\epsilon_{ij}\Phi_{2}^{\prime i }\Phi_{1}^{j} + h.c. ).
\end{eqnarray}

\subsection{Lepton sector}
\subsubsection{Charged leptons}
The most general superpotential allowed by gauge invariance for charged leptons is given by:

\begin{align}
     W_{L/E}&=  -\hat{\ell}_{L}^{p}\hat{\Phi}'_{2}{h}_{2e}^{p\mu}\hat{e}_{L}^{\mu\; c}
    - \hat{\ell}_{L}^{\tau}\hat{\Phi}'_{2}{h}_{2e}^{\tau r}\hat{e}_{L}^{r\; c} 
    - \hat{\ell}_{L}^{p}\hat{\Phi}'_{1}{h}_{1E}^{p}\hat{E}_{L}^{c}+ \hat{E}_{L}\hat{\chi}'{g}_{\chi' E}\hat{E}_{L}^{c}  \nonumber \\
    &- \hat{E}_{L}\mu_{E}\hat{\mathcal{E}}_{L}^{c} + \hat{\mathcal{E}}_{L}\hat{\chi}g_{\chi\mathcal{E}}\hat{\mathcal{E}}_{L}^{c}  - \hat{\mathcal{E}}_L\mu_{\mathcal{E}}\hat{E}_{L}^{c} + \hat{E}_{L}\hat{\sigma} h_{\sigma e}^{E r}\hat{e}_{L}^{c\; r} + \hat{\mathcal{E}}_{L}\hat{\sigma}^{\prime}h_{\sigma' e}^{\mathcal{E} \mu}\hat{e}_{L}^{\mu\; c}
\end{align}

\noindent
where $p=e,\mu$ labels the first and second generation lepton doublets and $r=e,\tau$ is the index of the right handed charged leptons. Then, spontaneous symmetry breaking (SSB) leads to the mass matrix structure in the flavor basis $(e^{e},e^{\mu} , e^{\tau}, E, \mathcal{E})$:

\begin{align}
    \mathcal{M}_{E}&=\frac{1}{\sqrt{2}}\left(\begin{array}{ c c c |c c}
    v_{2}\Sigma_{11}                           & h_{2e}^{e\mu}v_{2}     & v_{2}\Sigma_{13} &  h_{1e}^{E}v_{1}    & 0 \\
    0                           & h_{2e}^{\mu\mu}v_{2}   & 0 &  h_{1\mu}^{E}v_{1}  & 0 \\
    h_{2e}^{\tau e}v_{2}  & 0                            & h_{2e}^{\tau\tau}v_{2} & 0 & 0 \\ \hline
    0 & 0 & 0 & {g}_{\chi E}v_{\chi} & -\mu_{E} \\
    0 & 0 & 0 & -\mu_{\mathcal{E}} & g_{\chi\mathcal{E}}v_{\chi}  \\
    \end{array} \right).
\end{align}

\begin{figure}[H]
    \centering
    \includegraphics[scale=0.23]{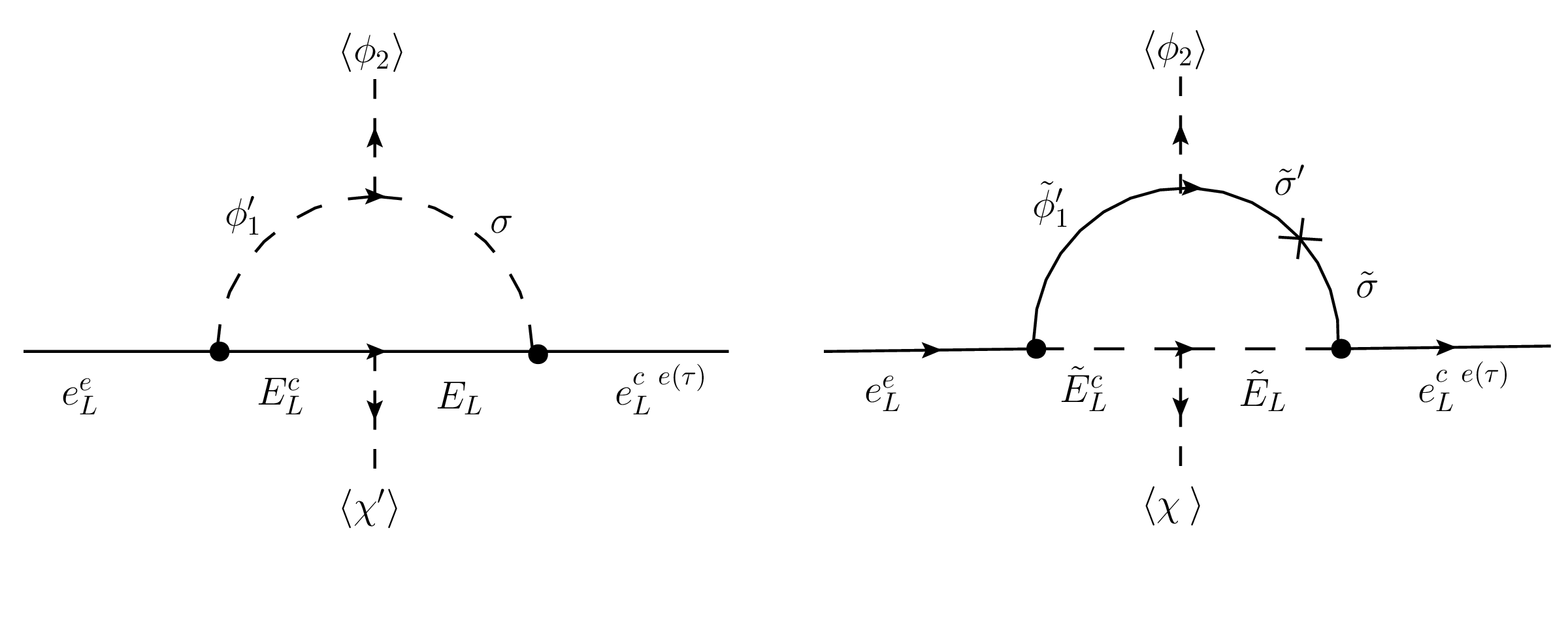}
    \caption{One loop corrections to the charged leptons mass matrix.}
    \label{fig:1-loopforleptons}
\end{figure}

Exotic leptons are expected to be highly massive which can be explained easily by a $U(1)_{X}$ symmetry breaking at a higher energy than the electroweak scale. In such case, exotic leptons are decoupled from SM leptons by a seesaw rotation. The $2\times 2$ submatrix for $\{E,\mathcal{E}\}$ exotic leptons can be diagonalized to eigenstates by an angle  $\theta_{E\mathcal{E}}^{L/R}$.  Moreover, the decoupled $3\times 3$ submatrix containing only SM leptons $\{e^{e},e^{\mu},e^{\tau}\}$ represents only two massive states, since the squared mass matrix $\mathcal{M}_{E}\mathcal{M}_{E}^{\dagger}$ has rank four, making the electron massless at tree level. Nevertheless, $\sigma$ and $\sigma'$ scalars mediate one-loop diagrams, shown in figure \ref{fig:1-loopforleptons}, making the electron massive by adding the following terms to the mass matrix:

\begin{align}
    \Delta \mathcal{L}_{L}&=\frac{v_{2}}{2}\left(\Sigma_{11}e^{e}_{L}e^{e}_{R} + \Sigma_{13}e^{e}_{L}e^{\tau}_{R}\right),
\end{align}

\noindent
The non-SUSY contribution is given by:

\begin{align}
    v_{2}\Sigma _{11(13)}^{NS}=\frac{-1}{16\pi ^2}\frac{v_{2}}{\sqrt{2}}\frac{\lambda_{1}\mu_{\sigma}h_{\sigma}^{e^c e(\tau)}h_{1E}^{e}}{M_{E}}C_0\left(\frac{m_{h1}^{\prime}}{M_{E}},\frac{m_{\sigma}^{\prime}}{M_E}\right).
\end{align}
\noindent

\noindent
where $M_{E}$ is the exotic charged fermion mass, $m_{h1}^{\prime}$ and $m_{\sigma}^{\prime}$ are the corresponding masses of the $h'_{1}$ and $\sigma'$ scalar fields in flavor basis respectively, $C_{0}$ is the Veltmann-Passarino function evaluated at $p^{2}=0$ given in Eq. (\ref{C0}). Whereas, the SUSY contribution is given by:
\small
\begin{align}
    v_{2}\Sigma_{11(13)}^{S}&(p^{2}=0) = -\frac{1}{32\pi^{2}}\frac{v_{2}}{\sqrt{2}}\sum_{n=1}^{10}\sum_{k=1}^{2}Z_{L}^{9n}Z_{L}^{4n} Z_{\tilde{h}}^{10 k}Z_{\tilde{h}}^{11 k} \lambda_{1}\mu_{\sigma}h_{\sigma}^{e^c e(\tau)}h_{1E}^{e}\times \\
    &\times\left[\frac{(\tilde{m}_{\sigma  k}+\tilde{m}_{h_{1}}^{\prime})^{2}}{\tilde{M}_{L_{n}}^{2}}C_{0}\left(\frac{\tilde{m}_{h1}^{\prime}}{\tilde{M}_{L_{n}}},\frac{\tilde{m}_{\sigma  k}}{\tilde{M}_{L_{n}}} \right) + \tilde{m}_{h1}^{\prime 2}B_{0}(0,\tilde{m}_{\sigma}^{\prime},\tilde{M}_{L_{n}}) + \tilde{m}_{\sigma k}^{2}B_{0}(0,\tilde{m}_{h1}^{\prime},\tilde{M}_{L_{n}}) \right] \nonumber
\end{align}

\begin{align}\label{C0}
    C_{0}(\hat{m}_{1},\hat{m}_{2}) &= \frac{1}{(1-\hat{m}_{1}^{2})(1-\hat{m}_{2}^{2})(m_{1}^{2}-\hat{m}_{2}^{2})}\left[\hat{m}_{1}^{2}\hat{m}_{2}^{2}Ln\left(\frac{\hat{m}_{1}^{2}}{\hat{m}_{2}^{2}}\right) + \hat{m}_{2}^{2}Ln(\hat{m}_{2}^{2})- \hat{m}_{1}^{2}Ln(\hat{m}_{1}^{2})\right],
\end{align}

\normalsize
\noindent
where $\tilde{M}_{L_{n}}$ are the charged sleptons mass eigenvalues, $Z_{\tilde{h}}$ is the rotation matrix that connects $\tilde{\sigma}$ ($\tilde{\sigma'}$) to their mass eigenstates, with mass eigenvalues $\tilde{m}_{h k}$, running into the loop. Likewise, $Z_{L}$ is the rotation matrix for exotic sleptons mass eigenstates $\tilde{L}_{n}$ and finally mass terms without an index are in flavor basis.

Mass eigenvalues and rotations of left handed leptons were obtained by diagonalizing the squared mass matrix $\mathcal{M}_{E}\mathcal{M}_{E}^{\dagger}$ which are obtained straightforwardly after the seesaw decoupling of exotic leptons masses. The final $2\times 2$ submatrix containing the electron an muon is diagonalized by a rotation angle $\theta_{e\mu}$ shown in Eq. (\ref{thetaemu}). As a result, mass eigenvalues are given by:
\begin{align}
m_{e}^{2}&=\frac{1}{2}v_{2}^{2}v_{2}^{2}\frac{t_{3}^{2}}{2m_{\tau}^{2}}, & m_{\mu}^{2}&=\frac{1}{2}v_2^{2}\left[(h_{2e}^{e\mu})^2+(h_{2e}^{\mu\mu})^2\right], \\
m_{\tau}^{2}&=\frac{1}{2}v_2^2\left[(h_{2e}^{\tau e})^2+(h_{2e}^{\tau \tau})^2\right], & m_{E}^{2}&=\frac{1}{2}g_{\chi' E}^2\; v_\chi^2, \\
m_{\mathcal{E}}^{2}&=\frac{1}{2}g_{\chi \mathcal{E}}^2 v_\chi^2.
 \end{align}

\noindent
The rotation matrix is written as the product of three matrices, $V^{L}=V_{3}^{\ell}V_{2}^{\ell}V_{1}^{\ell}$, which are given by:
\begin{align}\label{v1l}
  V_{1}^{\ell}&=\begin{pmatrix}
      1 & 0 & 0 & -\frac{g_{\chi \mathcal{E} } h_e^e v_{\chi } v'_1}{g_{\text{$\chi $E}} g_{\chi \mathcal{E} } v_{\chi } v'_{\chi }-\mu_{E} \mu _{\mathcal{E} }} & \frac{h_e^e \mu_{E} v'_1}{\mu_{E} \mu _{\mathcal{E} }-g_{\text{$\chi $E}} g_{\chi \mathcal{E} } v_{\chi } v'_{\chi }} \\
      0 & 1 & 0 & - \frac{g_{\chi \mathcal{E} } h_{\mu }^e v_{\chi } v'_1}{g_{\text{$\chi $E}} g_{\chi \mathcal{E} } v_{\chi } v'_{\chi }-\mu_{E} \mu _{\mathcal{E} }} & \frac{h_{\mu }^e \mu_{E} v'_1}{\mu_{E} \mu _{\mathcal{E} }-g_{\text{$\chi $E}} g_{\chi \mathcal{E} } v_{\chi } v'_{\chi }} \\
      0 & 0 & 1 & 0 &0 \\
      \frac{g_{\chi \mathcal{E} } h_e^e v_{\chi } v'_1}{g_{\text{$\chi $E}} g_{\chi \mathcal{E} } v_{\chi } v'_{\chi }-\mu_{E} \mu _{\mathcal{E} }}  &  \frac{g_{\chi \mathcal{E} } h_{\mu }^e v_{\chi } v'_1}{g_{\text{$\chi $E}} g_{\chi \mathcal{E} } v_{\chi } v'_{\chi }-\mu_{E} \mu _{\mathcal{E} }} & 0 & 1 &0 \\
       -\frac{h_e^e \mu_{E} v'_1}{\mu_{E} \mu _{\mathcal{E} }-g_{\text{$\chi $E}} g_{\chi \mathcal{E} } v_{\chi } v'_{\chi }} & -\frac{h_{\mu }^e \mu_{E} v'_1}{\mu_{E} \mu _{\mathcal{E} }-g_{\text{$\chi $E}} g_{\chi \mathcal{E} } v_{\chi } v'_{\chi }} & 0 & 0 &1 
    \end{pmatrix} ,
    \end{align}
    \begin{align}
    V_{2}^{\ell}&=\begin{pmatrix}
      1 & 0 & - \frac{m_{e}^{2}}{t_{3}v_{2}v_{2}^{\prime}} &0 & 0\\
      0 & 1 & 0 &0 & 0 \\
       \frac{m_{e}^{2}}{t_{3}v_{2}v_{2}^{\prime}} & 0 & 1 & 0 &0 \\
      0 & 0 & 0 & \cos\theta_{E\mathcal{E}}^{L} & -\sin\theta_{E\mathcal{E}}^{L} \\
      0 & 0 & 0 & \sin\theta_{E\mathcal{E}}^{L} & \cos\theta_{E\mathcal{E}}^{L} 
    \end{pmatrix}, & V_{3}^{\ell}&= \begin{pmatrix}
      \cos\theta_{e\mu} & \sin\theta_{e\mu} &0 &0 & 0\\
      -\sin\theta_{e\mu} & \cos\theta_{e\mu} &0 & 0 &0 \\
      0 & 0 & 1 & 0 & 0\\
      0 & 0 & 0 & 1 & 0 \\
      0 & 0 & 0 & 0 & 1
    \end{pmatrix} 
    \end{align}
    
\noindent
where $V_{1}^{\ell}$ decouples SM and exotic leptons, $V_{2}^{\ell}$ diagonalizes the exotic leptons submatrix and decouples the $\tau$ lepton, and $V_{3}^{\ell}$ allows to find the lightest eigenstates $\{e,\mu\}$; being $t_{3}$ and $\theta_{e\mu}$ parameters defined as:

    \begin{align}
    t_{3}&=\Sigma_{11}h_{2 e}^{\tau e}+\Sigma_{13} h_{2 e}^{\tau \tau },  &  \sin\theta_{e\mu}&=-\frac{2 m_{\mu}^{2}-v_2^{\prime 2} (h_{2 e}^{\mu\mu})^{2} }{ h_{2 e}^{e\mu } h_{2 e}^{\mu\mu} v_2^{\prime 2}\sqrt{1+\left(\frac{v_2^{\prime 2} (h_{2 e}^{\mu\mu})^{2} -2 m_{\mu}^{2}}{h_{2 e}^{e\mu  }  h_{2 e}^{\mu\mu } v_2^{\prime 2}}\right)^2}}. \label{thetaemu}
\end{align}

Likewise, rotation of right handed fermions come from the diagonalization of  $\mathcal{M}_{E}^{\dagger}\mathcal{M}_{E}$ which can be written as $V^{R}=U_{2}^{\ell}U_{1}^{\ell}$, where $U_{1}^{\ell}$ decouples the exotic leptons and $U_{2}^{\ell}$ diagonalizes exotic leptons by an angle $\theta_{E\mathcal{E}}^{R}$, and diagonalizes SM leptons by decoupling the muon and rotating the resulting $e-\tau$ mixing by an angle $\theta_{e\tau}$. Such rotations are given by:

\begin{align}
    U_{2}^{\ell}&=\begin{pmatrix}
 \mbox{\footnotesize $\cos\theta_{e\tau}$} & \frac{v_{2}^{2}h_{2 e}^{e\mu} \left(\sin\theta_{e\tau} \Sigma_{13}-\cos\theta_{e\tau} \Sigma_{11}\right) }{2 m_{\mu}^{2}} & \mbox{\footnotesize $-\sin\theta_{e\tau}$} & 0 & 0 \\
 \frac{\Sigma_{11} v_{2}^{2} h_{2 e}^{e\mu}}{2 m_{\mu}^{2}} & \mbox{\footnotesize $1$} & \frac{\Sigma_{13} v_{2}^{2} h_{2 e}^{e\mu}}{2 m_{\mu}^{2}} & 0 & 0 \\
 \mbox{\footnotesize $\sin\theta_{e\tau}$} & -\frac{v_{2}^{2} h_{2 e}^{e\mu} \left(\cos_{\theta_{e\tau}}\Sigma_{13}+\sin\theta_{e\tau} \Sigma_{11}\right) }{2 m_{\mu}^{2}} & \mbox{\footnotesize $\cos\theta_{e\tau}$} & 0 & 0 \\
 0 & 0 & 0 & \mbox{\footnotesize $\cos\theta_{E\mathcal{E}}^{R}$} & \mbox{\footnotesize $-\sin\theta_{E\mathcal{E}}^{R}$} \\
 0 & 0 & 0 & \mbox{\footnotesize $\sin\theta_{E\mathcal{E}}^{R}$} & \mbox{\footnotesize $\cos\theta_{E\mathcal{E}}^{R}$} \\
\end{pmatrix}, &
U_{1}^{\ell}&=\begin{pmatrix}
\mbox{\footnotesize $\mathcal{I}_{3\times 3}$} & \mbox{\footnotesize $-\Theta^{T}$}  \\
\mbox{\footnotesize $\Theta$} & \mbox{\footnotesize $\mathcal{I}_{2\times 2}$}
\end{pmatrix}
\end{align}
 with
 
\small
\begin{align}
    \Theta^{T}=\frac{v_{1}v_{2}}{4m_{E}^{2}m_{\mathcal{E}}^{2}}
    \begin{pmatrix}
        \Sigma_{11} h_{1e}^{E} \left(g_{\chi \mathcal{E} }^2 v_{\chi }^2+\mu_{E}^{2}\right) & \Sigma_{11} h_{1e}^{E} v_{\chi} \left(\mu_{E} g_{\chi E} +g_{\chi \mathcal{E} } \mu _{\mathcal{E}}\right)  \\
       \left(g_{\chi \mathcal{E} }^2 v_{\chi }^2+\mu_E^2\right) \left(h_{1e}^{E} h_{2e}^{e\mu}+h_{\mu }^{E} h_{2e}^{\mu \mu }\right)  & v_{\chi}\left(h_{1e}^{E} h_{2 e}^{e\mu}+h_{1\mu}^{E} h_{2 e}^{\mu \mu }\right) \left(\mu_{E} g_{\chi E}+g_{\chi \mathcal{E} } \mu_{\mathcal{E}}\right) \\
        \Sigma_{13} h_{1e}^{E} \left(g_{\chi \mathcal{E} }^2 v_{\chi }^2+\mu_E^2\right)         
  & \Sigma_{13} h_{1e}^E v_{\chi} \left(\mu_E g_{\chi E} +g_{\chi \mathcal{E} }\mu_{\mathcal{E} }\right) 
    \end{pmatrix}.
\end{align}

\normalsize

\subsubsection{Neutral leptons}
Now, the neutrino superpotential is given by:
\begin{align}
            W_{L/N}&=\hat{\ell}_{L}^{p}\hat{\Phi}_{2}h_{2\nu}^{pq}\hat{\nu}_{L}^{q\; c}+\hat{\nu}_{L}^{q\; c}\hat{\chi}' {h}_{\chi N}^{\prime\; qn}\hat{N}_{L}^{n\; c}
        + \frac{1}{2}\hat{N}_{L}^{m\; c} M_{mn}\hat{N}_{L}^{n\; c}
\end{align}

\noindent
where $p=e,\mu$, $q=e,\mu,\tau$ labels the right handed neutrinos,  and $m,n=1,2,3$ label the Majorana neutrinos. After SSB, the $9\times 9$ mass matrix arises in the  basis $\left(\begin{matrix}{\nu^{e,\mu,\tau}_{L}},\,\left(\nu^{e,\mu,\tau}_{R}\right)^{C},\,\left(N^{e,\mu,\tau}_{R}\right)^{C}\end{matrix}\right)$, given by:

\begin{align}
\mathcal{M}_{\nu} &=\begin{pmatrix}
    0 & m_{D} & 0 \\
    m_{D}^{T} & 0 & M_{D} \\
    0 & M_{D}^{T} & M_{M}
    \end{pmatrix},     
\end{align}

where the block matrices are defined as:
\begin{align}
 &m_{D}=\frac{v_{2}}{\sqrt{2}}\begin{pmatrix}
    h_{2e}^{\nu e} & h_{2e}^{\nu \mu} & h_{2e}^{\nu \tau} \\
    h_{2\mu}^{\nu e} & h_{2\mu}^{\nu \mu} & h_{2\mu}^{\nu \tau} \\
    0 & 0 & 0
    \end{pmatrix},\ \ \
     (M_{D})^{ij}=\frac{v'_{\chi}}{\sqrt{2}}({h}_{\chi}^{\prime \nu})^{ij}, \ \ \ \ \  (M_{M})_{ij}=\frac{1}{2}M_{ij}.
\end{align}

To generate neutrino masses via inverse seesaw mechanism the hierarchy $M_{M}\ll m_{D} \ll M_{D} $ is assumed and block diagonalization is achieved by the $\mathbb{V}_{SS}$ matrix given by:

\begin{align}
\mathbb{V}_{SS}\mathcal{M}_{\nu}\mathbb{V}_{SS}^{\dagger}&\approx\begin{pmatrix}
m_{light}&0\\
0&m_{heavy}
\end{pmatrix}, &  \\
\mathbb{V}_{SS}&=\begin{pmatrix}
I&-\Theta_{\nu}\\
\Theta_{\nu}^{T}&I
\end{pmatrix}, & \Theta_{\nu}&=\begin{pmatrix}
0&M_{D}^{T}\\
M_{D}&M_{M}
\end{pmatrix}^{-1}\begin{pmatrix}
m_{D}^{T}\\0
\end{pmatrix},
\end{align}

\noindent
where $m_{light}=m_{D}^{T}(M_{D}^{T})^{-1}M_{M}(M_{D})^{-1}m_{D}$ is the $3\times3$ mass matrix containing the active neutrinos and $m_{heavy}$ in Eq. (\ref{nuheavy}) contains six heavy Majorana neutrino mass eigenstates:
\begin{align}\label{nuheavy}
M_{D} &= \frac{v_{\chi}}{\sqrt{2}} \left( \begin{matrix}
h_{N\chi e}	&	0	&	0	\\	0	&	h_{N\chi \mu}	&	0	\\	0	&	0	&	h_{\chi N \tau}
\end{matrix} \right) &
M_{M} &= \mu_{N} \mathbb{I}_{3\times 3} & m_{heavy}\approx\begin{pmatrix}0&M_{D}^{T}\\
M_{D}&M_{M}
\end{pmatrix}  .  
\end{align}
For simplicity we consider the particular case of $M_{D}$ being  diagonal and $M_{M}$ proportional to the identity. Thus, light neutrino mass matrix takes the form:

\begin{equation}\label{mnu}
m_{\mathrm{light}} = \frac{\mu_{N} v_{2}^{2}}{{h_{N\chi e}}^{2}v_{\chi}^{2}}
\left( 
\begin{matrix}
	\left( h_{2e}^{\nu e}\right)^{2} + \left( h_{2\mu}^{\nu e} \right)^{2} \rho^{2} &
	{h_{2e}^{\nu e}}\,{h_{2e}^{\nu \mu}} + {h_{2\mu}^{\nu e}}\,{h_{2\mu}^{\nu \mu}}\rho^2 	&
	{h_{2e}^{\nu e}}\,{h_{2e}^{\nu \tau}}+ {h_{2\mu}^{\nu e}}\,{h_{2\mu}^{\nu \tau}}\rho^2 	\\
	{h_{2e}^{\nu e}}\,{h_{2e}^{\nu \mu}} + {h_{2\mu}^{\nu e}}\,{h_{2\mu}^{\nu \mu}}\rho^2	&	
	\left( h_{2e}^{\nu \mu} \right)^{2} + \left( h_{2\mu}^{\nu\mu} \right)^{2} \rho^{2}	&	
	{h_{2e}^{\nu \mu}}\,{h_{2e}^{\nu \tau}}+ {h_{2\mu}^{\nu \mu}}\,{h_{2\mu}^{\nu \tau}}\rho^2	\\
	{h_{2e}^{\nu e}}  \,{h_{2e}^{\nu \tau}}+ {h_{2\mu}^{\nu e}}  \,{h_{2\mu}^{\nu \tau}}\rho^2	&	
	{h_{2e}^{\nu \mu}}\,{h_{2e}^{\nu \tau}}+ {h_{2\mu}^{\nu \mu}}\,{h_{2\mu}^{\nu \tau}}\rho^2	&	
	\left( h_{2e}^{\nu \tau} \right)^{2} + \left( h_{2\mu}^{\nu \tau} \right)^{2} \rho^{2}
\end{matrix} \right),
\end{equation}

\noindent
where $\rho={h_{N\chi e}}/{h_{N\chi \mu}}$. Similarly,  $m_{\mathrm{light}}$ contains a single massless neutrino although such possibility is still allowed because we know from experiments only squared mass differences. Besides, exotic neutrinos, mass eigenstates can be obtained easily from Eq. (\ref{nuheavy}) and are labeled as $\mathcal{N}^{k}$, $k=1,...,6.$, which can be read as:
\begin{align}
    m_{\mathcal{N}^{1}}&=\frac{1}{2}(\mu_{N}-\sqrt{\mu_{N}^{2}+2h_{N_{\chi 1}}v_{\chi}^{2}}) & m_{\mathcal{N}^{2}}&=\frac{1}{2}(\mu_{N}-\sqrt{\mu_{N}^{2}+2h_{N_{\chi 2}}v_{\chi}^{2}}) \\
    m_{\mathcal{N}^{3}}&=\frac{1}{2}(\mu_{N}+\sqrt{\mu_{N}^{2}+2h_{N_{\chi 1}}v_{\chi}^{2}}) & m_{\mathcal{N}^{4}}&=\frac{1}{2}(\mu_{N}+\sqrt{\mu_{N}^{2}+2h_{N_{\chi 2}}v_{\chi}^{2}}) \\
    m_{\mathcal{N}^{5}}&=\frac{1}{2}(\mu_{N}-\sqrt{\mu_{N}^{2}+2h_{N_{\chi 3}}v_{\chi}^{2}}) & m_{\mathcal{N}^{6}}&=\frac{1}{2}(\mu_{N}+\sqrt{\mu_{N}^{2}+2h_{N_{\chi 3}}v_{\chi}^{2}})     
\end{align}

\section{Electron and muon EDM}
\subsection{One-loop contribution}
\begin{figure}
    \centering
    \includegraphics[scale=0.6]{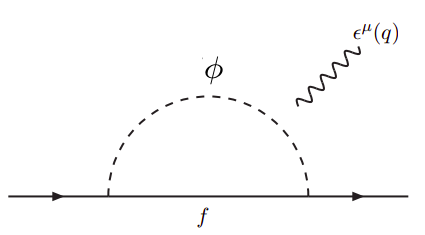}
    \caption{1-loop contributions to EDM by charged leptons. The photon external leg can be attached to the scalar or fermion internal line, $\phi=H_{i}^{+},\tilde{\nu}$ being $i=2,3,4$ and $f=\mathcal{N}_{j}, \tilde{\chi}^{\pm}$ with $j=1,...,6$.}
    \label{1loop-barzee}
\end{figure}

Despite the SM prediction of the EDM is considerably small, interactions with scalar particles may add significant contributions at one-loop and two-loop level. In this model, additional CP violation comes from exotic neutrino contributions, which at one-loop level contribute via the diagram shown in figure \ref{1loop-barzee}. The contribution is given by \cite{1loopformula1}\cite{1loopformula2}:

\begin{align}
    d_{e}&=\frac{m_{f}}{16\pi^{2}m_{\phi}^{2}}Im[Y_{f \phi}^{L}Y_{f \phi}^{R *}]\left[Q_{f}A\left(\frac{m_{f}^{2}}{m_{\phi}^{2}}\right) + Q_{\phi}B\left(\frac{m_{f}^{2}}{m_{\phi}^{2}}\right)\right]
\end{align}

\noindent
where $Q_{f}$ and $Q_{\phi}$ represent the electric charge of the fermion $f$ and scalar $\phi$ respectively and $Y_{f\phi}^{L/R}$ are Yukawa couplings related to the interaction lagrangian given by:
\begin{align}
    \mathcal{L}(f,\phi)&=Y_{f\phi}^{L}\bar{\psi}_{f}P_{L}\psi_{e(\mu)}\phi + Y_{f\phi}^{R}\bar{\psi}_{f}P_{R}\psi_{e(\mu)}\phi, 
\end{align}   

\noindent
where $\psi_{f}$ represents the fermion running into the loop and $\psi_{e(\mu)}$ the electron (muon). Besides, $A$ and $B$ are loop functions given by:
\begin{align}
    A(r)&=\frac{1}{2(1-r)^{2}}\left(3-r-\frac{2 ln(r)}{1-r}\right) & B(r)&=\frac{1}{2(1-r)^{2}}\left(1+r+\frac{2r ln(r)}{1-r}\right).
\end{align}
    
\begin{figure}[H]
    \centering
    \includegraphics[scale=0.51]{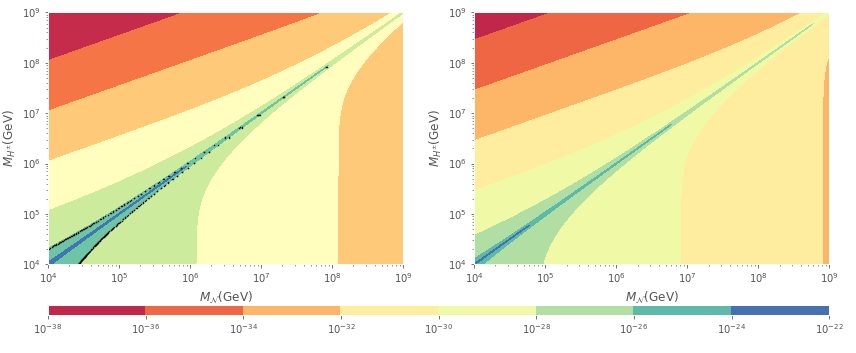}
    \caption{One-loop absolute value contribution to electron (left) and muon (right) EDM due to the lightest exotic neutrino interacting with the charged scalar $H_{3}$. The black dots represents the current experimental upper bound and the contribution is negative for all masses.}
    \label{1loop}
\end{figure}

First, lets consider the contributions due to charged scalars and exotic neutrinos ($\phi=H^{+}$, $f=\mathcal{N}$) whose interaction lagrangian in mass basis is given by:    
    
\begin{align}
  \mathcal{L}(\mathcal{N},H^{+}) &\approx  \bar{\mathcal{N}}^{j} H_{i}^{+}\Bigg[ - P_{L} \frac{R^{H^{+}}_{i,3}}{\sqrt{2} }  \left(c_{\theta_{e\mu}} h_{2e}^{\nu k }+s_{\theta_{e \mu}} h_{2\mu}^{\nu k }\right)e^{e} - P_{L} \frac{R^{H^{+}}_{i,3}}{\sqrt{2} }  \left(-s_{\theta_{e\mu}} h_{2e}^{\nu k }+c_{\theta_{e \mu}} h_{2\mu}^{\nu k }\right)e^{\mu} \nonumber\\
  & +P_{R} \left(R_{3+j,1}^{\nu } (R^{H^{+}}_{2,i} V_{p,4}^{R} h_{1e}^{E}+R^{H^{+}}_{4,i} V_{p,2}^{R} h_{2e}^{e\mu}) +R_{3+j,2}^{\nu } (R^{H^{+}}_{2,i} V_{p,4}^{R} h_{1\mu}^{E}+R^{H^{+}}_{4,i} V_{p,2}^{R} h_{2e}^{\mu\mu})\right) e^{p} \bigg],
 \end{align}
  
  \noindent
  where $p=e, \mu$ labels the external fermion, $i=2,3,4$ sums over the three charged scalar field mass eigenstates, $j=1,2,3,4,5,6$ labels the exotic heavy neutrino eigenstates and $k$ is an index dependent on $j$ to label the neutrino Yukawa couplings, defined as $k(1)=k(3)=e$, $k(2)=k(4)=\mu$ and $k(4)=k(5)=\tau$. Nonetheless,  $R^{H^{+}}$ and $V^{R}$ are the rotation matrices for charged scalars and right-handed leptons respectively, whereas $R^{\nu}$ is the $9\times 9$ rotation matrix for neutrinos.
  On the one hand, after getting the couplings numerically we have found that in addition to the dependence of the second scalar mass eigenstate on $v_{\chi}$ and $v'_{\chi}$,  its coupling is inversely dependent on $v_{\chi}$ and $v'_{\chi}$ because of the rotation matrix, so the EDM contribution becomes highly suppressed by $H_{2}^{+}$ mass. On the other hand, the remaining two heavy eigenstates masses depend on the free soft SUSY breaking parameters $\mu_{11}$ and $\mu_{22}$ for $H_{3}^{+}$ and $H_{4}^{+}$ respectively. Thus, we can vary $H_{3}^{+}$, $H_{4}^{+}$ and $\mathcal{N}$ masses independently without suppressing the coupling for large masses, leading to the contributions shown in figure \ref{1loop}. All couplings between exotic neutrinos and charged scalars are of order $\sim 10^{-1}$ so when adding all possibilities in the loop, the final EDM prediction differs from the values of figure \ref{1loop} at most, by a factor of 10.

\noindent

\begin{figure}[H]
    \centering
    \includegraphics[scale=0.55]{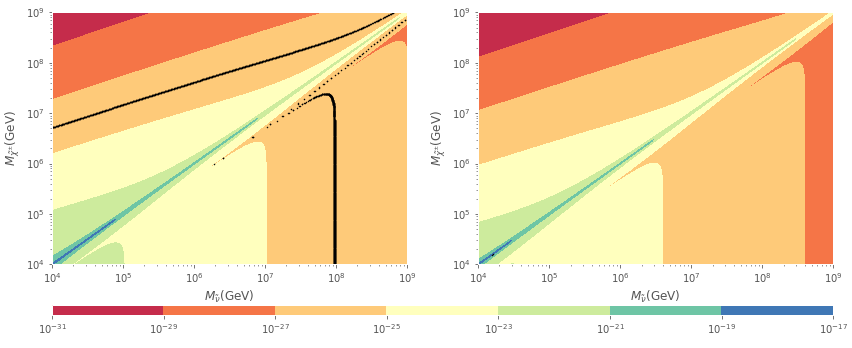}
    \caption{One-loop absolute value contribution to electron (left) and muon (right) EDM due to sneutrinos interacting with charginos. The black dots represents the current experimental upper bound. The upper (lower) triangular subsection separated by the $M_{\tilde{\nu}}=M_{\tilde{\chi}^{\pm}}$ line for the electron (muon) contribution is actually negative.}
    \label{1loopSUSY}
\end{figure}

Likewise, supersymmetry makes sneutrinos to have CP violating  complex couplings as well which leads to similar contributions to EDM according to the diagram in figure \ref{1loop-barzee}, the associated interaction lagrangian is given by:

  \begin{align}
    \mathcal{L}(\chi^{+},\tilde{\nu})&=\bar{\tilde{\chi}}^{+}_{i} P_{L}  \Bigg[g (R^{\tilde{\nu}}_{j,1} c_{\theta_{e\mu}} + R^{\tilde{\nu}}_{j,2} s_{\theta_{e\mu}} ) R^{\tilde{\chi}^{+}}_{i,1}e^{e} + g (-R^{\tilde{\nu}}_{j,1} s_{\theta_{e\mu}} + R^{\tilde{\nu}}_{j,2} c_{\theta_{e\mu}} ) R^{\tilde{\chi}^{+}}_{i,1}e^{\mu} \nonumber \\
    &-\Big((R^{\tilde{\nu}}_{4,j} h_{2e}^{\nu e} + R^{\tilde{\nu}}_{5,j} h_{2e}^{\nu \mu} + R^{\tilde{\nu}}_{6,j} h_{2e}^{\nu\tau}) c_{\theta_{e\mu}}e^{e} - (R^{\tilde{\nu}}_{4,j} h_{2e}^{\nu e} + R^{\tilde{\nu}}_{5,j} h_{2e}^{\nu \mu} + R^{\tilde{\nu}}_{6,j} h_{2e}^{\nu\tau}) s_{\theta_{e\mu}}e^{\mu} \nonumber \\
    &- (R^{\tilde{\nu}}_{4,j} h_{2\mu}^{\nu e} + R^{\tilde{\nu}}_{5,j} h_{2\mu}^{\nu\mu} + 
         R^{\tilde{\nu}}_{6,j} h_{2\mu}^{\nu\tau}) s_{\theta_{e\mu}}e^{e} - (R^{\tilde{\nu}}_{4,j} h_{2\mu}^{\nu e} + R^{\tilde{\nu}}_{5,j} h_{2\mu}^{\nu\mu} + 
         R^{\tilde{\nu}}_{6,j} h_{2\mu}^{\nu\tau}) c_{\theta_{e\mu}}e^{\mu} \Big) R^{\tilde{\chi}^{+}}_{i,3}\Bigg]\tilde{\nu}_{j} \nonumber\\
         &+\bar{\tilde{\chi}}^{+}_{i} P_{R}\Bigg[R^{\tilde{\chi}^{-}}_{i,3} (R^{\tilde{\nu}}_{j,1} h_{2e}^{e\mu} + R^{\tilde{\nu}}_{j,2} h_{2e}^{\mu\mu}) \Bigg]\tilde{\nu}_{j} e^{p}
\end{align}

\noindent
where $p=e,\mu$ labels the external fermion, $R^{\tilde{\nu}}$ and $R^{\tilde{\chi}^{\pm}}$ are the rotation matrices for sneutrinos and charginos and $g$ is the electroweak coupling constant. Since charginos and sneutrinos are expected to have big masses, soft breaking mass terms dominate sneutrino masses making electroweak contributions in mass matrices negligible, so mass matrices are approximately diagonal. As a result, we can change their masses independently of each other and the second chargino does not have important contributions. The result is shown in figure \ref{1loopSUSY} for the first chargino and sneutrino mass eigenstates.

In the case of the electron, big masses are required for SUSY particles in order to have an EDM contribution lying under the experimental limits. Besides, the model contains three charginos and nine sneutrinos that makes 27 possible interactions to be considered in each vertex. However, interactions with the lightest chargino have negligible small couplings ($\sim 10^{-9}$) and so as well many other interactions with other heavy charginos. Nevertheless, in figure \ref{1loopSUSY} shows the dominant contribution which is achieved by $\tilde{\chi}^{\pm}$ and $\tilde{\nu}_{8}$.

From both one-loop contributions it is clear that charged scalars and exotic neutrinos cannot have similar masses, and in a similar fashion for charginos and sneutrinos. Besides, the SUSY contributions tell us that if either a sneutrino mass is close to be experimentally measured ($<10^{6}$ GeV), charginos must have a greater mass ($\gtrapprox 10^{7}$ GeV) while if the chargino is near to observation, sneutrino would have  heavy masses, greater than $10^{8}$ GeV. However, if the chargino-sneutrino interaction is a new source of CP violation, it would imply a lower bound for the muon EDM since for both particles the contribution to EDM is similar.

\subsection{Two-loop contribution}
Due to the EDM smallness, two-loop contributions have been considered and it was initially found by Barr and Zee \cite{barr-zee} that there are several two-loop diagrams with important contributions to fermion EDM because the heavy internal fermion makes the contribution proportional to its mass. Besides, an additional CP violation source can come from any particle coupling since we are dealing with a huge amount of particles and free parameters. Contributions due to charginos and neutralinos have been considered in \cite{charginos} as well as the gluonic dimension-6 Weinberg operator \cite{gluonic} and CP-odd four-fermion operators \cite{fourfermion}. Moreover, effects of squarks have been studied in \cite{ellis} but in this model we focus on the effects of exotic neutrinos and sneutrinos in an inverse seesaw mechanism for neutrino mass generation.



\begin{figure}[H]
\centering
\includegraphics[scale=0.1]{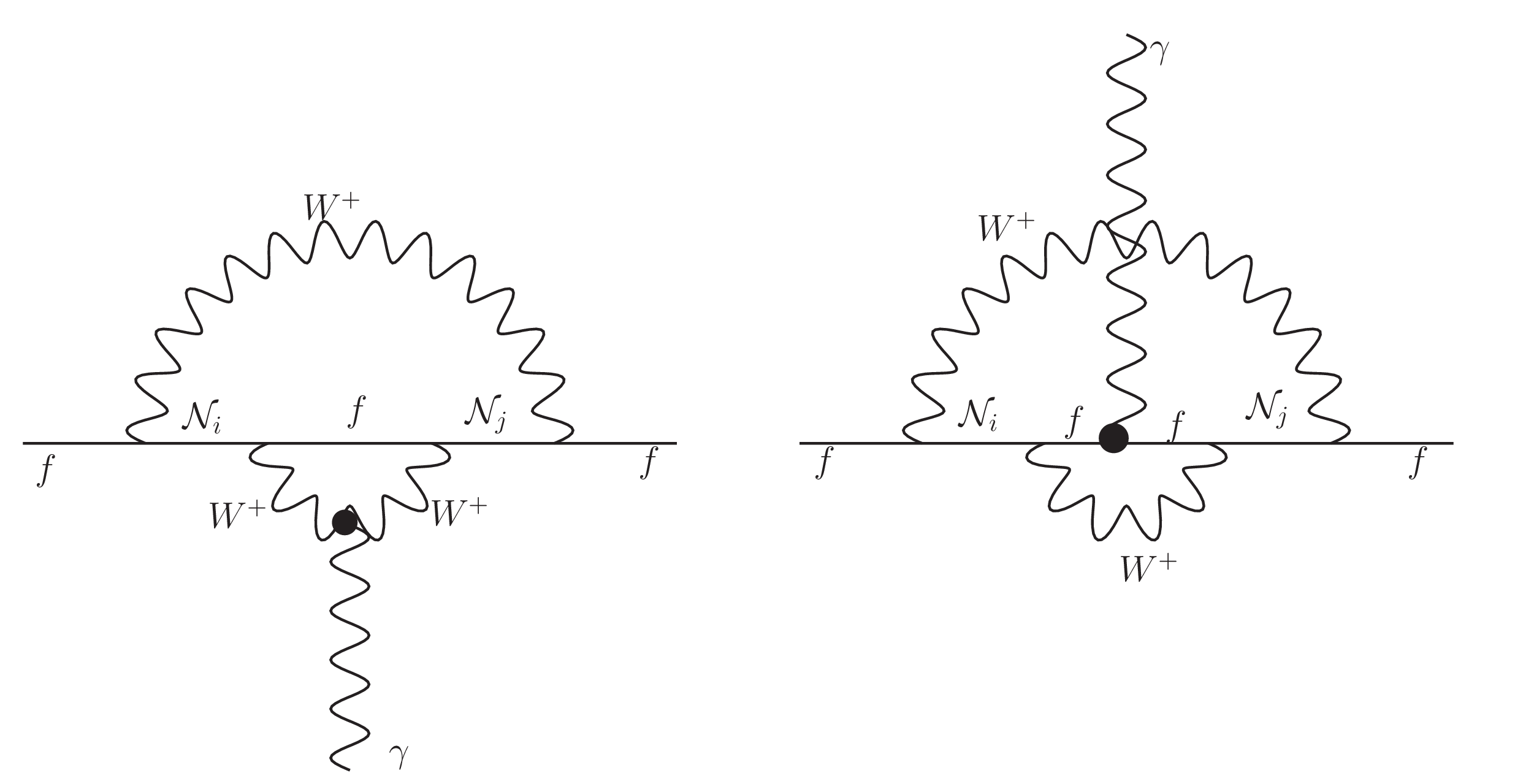}
\caption{Two-loop contribution to charged leptons EDM due to $W$ bosons.}
\label{2loopW}
\end{figure}

First, the contribution due to $W$ bosons have already been studied in \cite{asmaISS}\cite{asmasterileneutrinos} where they find that the main contribution comes from the heavy (sterile) neutrinos which provide dominant contributions from pseudo-Dirac pairs. The diagrams are shown in figure \ref{2loopW}. Accounting for PMNS unitarity and experimental bounds on sterile neutrinos decaying to a $W$ boson and a charged lepton they find that sterile neutrino masses have a mass upper bound given by:

\begin{align}
m_{i}\leq 873 GeV \left(\sum_{\alpha} |U_{\alpha u}|^{2}\right)^{-\frac{1}{2}}
\end{align}

\noindent
where $U_{\alpha i}$ is the extendend PMNS matrix with extra neutrinos. Nonetheless, the mass upper bound can be increased if we assume that exotic neutrinos dominant decay is to a charged scalar or SUSY particle yet unobserved. This is possible since the coupling of exotic neutrinos to charged leptons is suppresed by $v_{\chi}^{-1}$ as it can be seen in the lagrangian. The contribution to electron EDM is given by:
\begin{align}
d_{e} \approx -\frac{4}{3}\frac{g_{2}^{4}em_{e}}{4(4\pi)^{2}M_{W}^{2}}\sum_{\beta}\sum_{i, j}\left[J_{ije\beta}^{M}\mathcal{I}_{M}\left(\frac{m_{i}}{m_{W}},\frac{m_{j}}{m_{W}}\right) + J_{ije\beta}^{D}\mathcal{I}_{D}\left(\frac{m_{i}}{m_{W}},\frac{m_{j}}{m_{W}}\right) 	\right]
\end{align}

\noindent
where the $\sfrac{4}{3}$ comes from the consideration of three right-handed neutrinos and three sterile neutrinos, $J_{ij\alpha\beta}^{M}\equiv Im[U_{\alpha j}U_{\beta j}U_{\beta i}^{*}U_{\alpha i}^{*}]$ and $J_{ij\alpha\beta}^{D}\equiv Im[U_{\alpha j}U_{\beta j}^{*}U_{\beta i}U_{\alpha i}^{*}]$, $\mathcal{I}^{M}$ and $\mathcal{I}^{D}$ are loop functions that can be consulted in the appendix of ref. \citep{asmaISS}. It was shown that such contributions can be in agreement with electron EDM experimental upper bound for $J^{D}\sim 10^{-8}-10^{-12}$ or smaller in our case of heavier neutrinos. Likewise, it is possible due to the $v_{\chi}$ factor in the lagrangian which makes $J^{D} \sim 10^{-12}$ if we assume $\chi$ breaking scale at the order of the TeV scale. \\

\begin{figure}[H]
    \centering
    \includegraphics[scale=0.08]{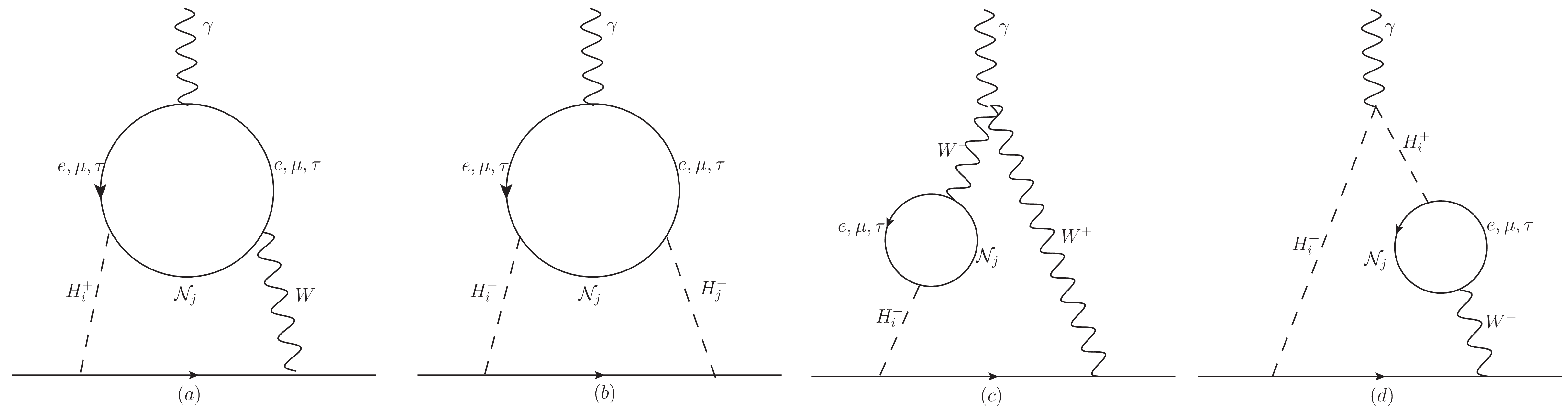}
    \caption{Two-loop Barr-Zee diagrams contributing to electron and muon EDM.}
    \label{2loop-barrzee}
\end{figure}

Now, our interest lies in the contributions due to charged scalars and SUSY particles in this inverse-seesaw scheme that generate additional contributions due to Barr-Zee diagrams shown in figure \ref{2loop-barrzee}.

\begin{figure}[H]
    \centering
    \includegraphics[scale=0.6]{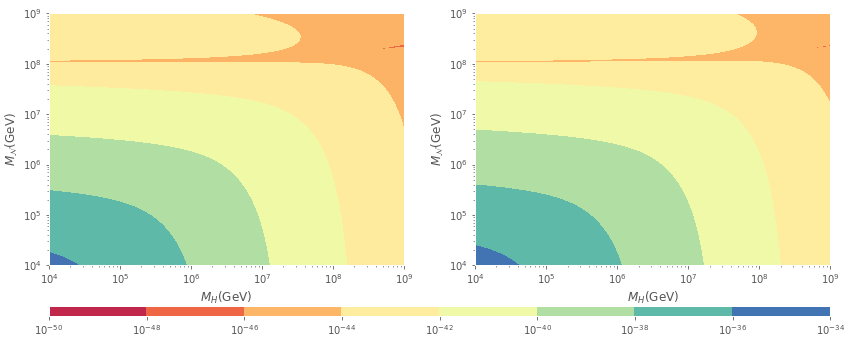}
    \caption{Absolute value of the contribution to electron (left) and muon (right) EDM due to charged scalars, $W$ boson and exotic neutrinos as a function of their masses for a $\tau$ lepton into the internal loop.}
    \label{2loopgraph}
\end{figure}

The contributions due to figure \ref{2loop-barrzee}a and \ref{2loop-barrzee}c is given by \cite{generalEDMformula}:
\begin{align}
d_f^{S^*V+SV^*} = -\frac{1}{32 \pi^2 m_H^2} & \int_0^1 dx \frac{1}{1-x} j\left(\frac{m_W^2}{m_H^2}, \frac{{\tilde{\Delta}}}{m_H^2}\right) \times  \nonumber \\
& \times \left[Im[(g_{Hf}^{R*}g_{Wf}^{L}+g_{Hf}^{L*}g_{Wf}^{R})c_{0}^{S^{*}V}]\right.  \left. +Im[(g_{Hf}^{R*}g_{Wf}^{R}-g_{Hf}^{L*}g_{Wf}^{L})c_{E}^{S^{*}V}]\right]. \label{dHW}
\end{align}

where the coefficients $\tilde{\Delta}$, $c_{E}^{S^{*}V}$ and $c_{0}^{S^{*}V}$ are related to the inner loop and given by:

\begin{align}
c_E^{S^* V} &= -\frac{e Q_f N_c}{8\pi^2} \left[m_{\mathcal{N}} (1-x)^2 \left(Y_{Hf}^{R}Y_{Wf}^{R*}+Y_{Hf}^{L}Y_{Wf}^{L*}\right) + x^2 m_f \left(Y_{Hf}^{R}Y_{Wf}^{L*}+Y_{Hf}^{L}Y_{Wf}^{R*}\right)\right], \nonumber \\
c_O^{S^* V} &= -\frac{e Q_f N_c }{8\pi^2} \left[m_{\mathcal{N}} (1-x) \left(Y_{Hf}^{R}Y_{Wf}^{R*}-Y_{Hf}^{L}Y_{Wf}^{L*}\right) + x m_f \left(Y_{Hf}^{R}Y_{Wf}^{L*}-Y_{Hf}^{L}Y_{Wf}^{R*}\right)\right],
\end{align}
\begin{align}
\tilde{\Delta}&= \frac{x m_{\mathcal{N}}^2 + (1-x) m_f^2}{x(1-x)} & j(r,s) &= \frac{1}{r-s} \left(\frac{r\log r}{r-1} - \frac{s\log s}{s-1}\right)
\end{align}
 being the $g_{HF}^{L(R)}$ and $g_{WF}^{L(R)}$ couplings related to the outer loop and the $Y_{HF}^{L(R)}$  and $Y_{WF}^{L(R)}$ couplings related to the inner loop. Moreover, such formula agrees with the presented in \cite{generalEDMformula} and \cite{MSSMcase} in the MSSM case when all $Y^{R}$ couplings in Eq. (\ref{dHW}) are zero. Furthermore, such diagrams require the interaction with $W$ bosons, whose interaction lagrangian is given by:
 
 \begin{align}
      \mathcal{L}(W,\mathcal{N})&=-\bar{\mathcal{N}}^{j} W_{\mu}^{+} \frac{g_w  v_{2} \gamma^{\mu}P_{L}}{2h_{N_{\chi k}} v_{\chi}} \left[ c_{\theta_{e\mu}} h_{2e}^{\nu k} +s_{\theta_{e\mu}}h_{2\mu}^{\nu k} \right]e^{e} \nonumber \\
      &\;\;\;\;\;\;\;\;\;\;\;\;\;\;\;\;\;\;\;\;\;\;\;\;\;\;\;\;\;\;\;\;\;\;\;\;\;\;\;\;\;\;\;\;\;\;\;-\bar{\mathcal{N}}^{j} W_{\mu}^{+} \frac{g_w  v_{2} \gamma^{\mu}P_{L}}{2h_{N_{\chi k}} v_{\chi}} \left[ -s_{\theta_{e\mu}} h_{2e}^{\nu k} +c_{\theta_{e\mu}}h_{2\mu}^{\nu k} \right] e^{\mu}
 \end{align}
where, similarly, $j=1,2,3,4,5,6$ labels the exotic heavy neutrino eigenstates and $k$ is an index dependent on $j$ to label the neutrino Yukawa couplings, defined as $k(1)=k(3)=e$, $k(2)=k(4)=\mu$ and $k(4)=k(5)=\tau$. In addition to diagram \ref{2loop-barrzee}a, diagrams shown in \ref{2loop-barrzee}c and \ref{2loop-barrzee}d are required to achieve a gauge invariant contribution and there are no diagrams with Goldstone bosons, that is because in the non-linear $R_{\xi}$ gauge they vanish as well as diagram \ref{2loop-barrzee}d does when $\xi\rightarrow 0$ because the internal loop is proportional to the four momentum of the $W$ boson and the $W$ propagator is transverse in Landau gauge \cite{MSSMcase}. The contribution as a function of the neutrino and charged scalar mass is shown in figure \ref{2loopgraph}\\

Finally, the contribution to EDM due to $H^{\pm}$ is shown in figure \ref{2loop-barrzee}b which is non zero but can be neglected as it corresponds to a loop insertion in the one-loop diagram shown in figure \ref{1loop-barzee}. Moreover, the same diagrams with gauge bosons instead of scalars provides a null contribution because $W$ bosons do not couple with right-handed charged leptons. However, it is non-zero if charginos and neutralinos run into the loop \cite{WWdiagram} but in our scenario there is no CP-violation sources in such interactions.



\begin{figure}[H]
    \centering
    \includegraphics[scale=0.15]{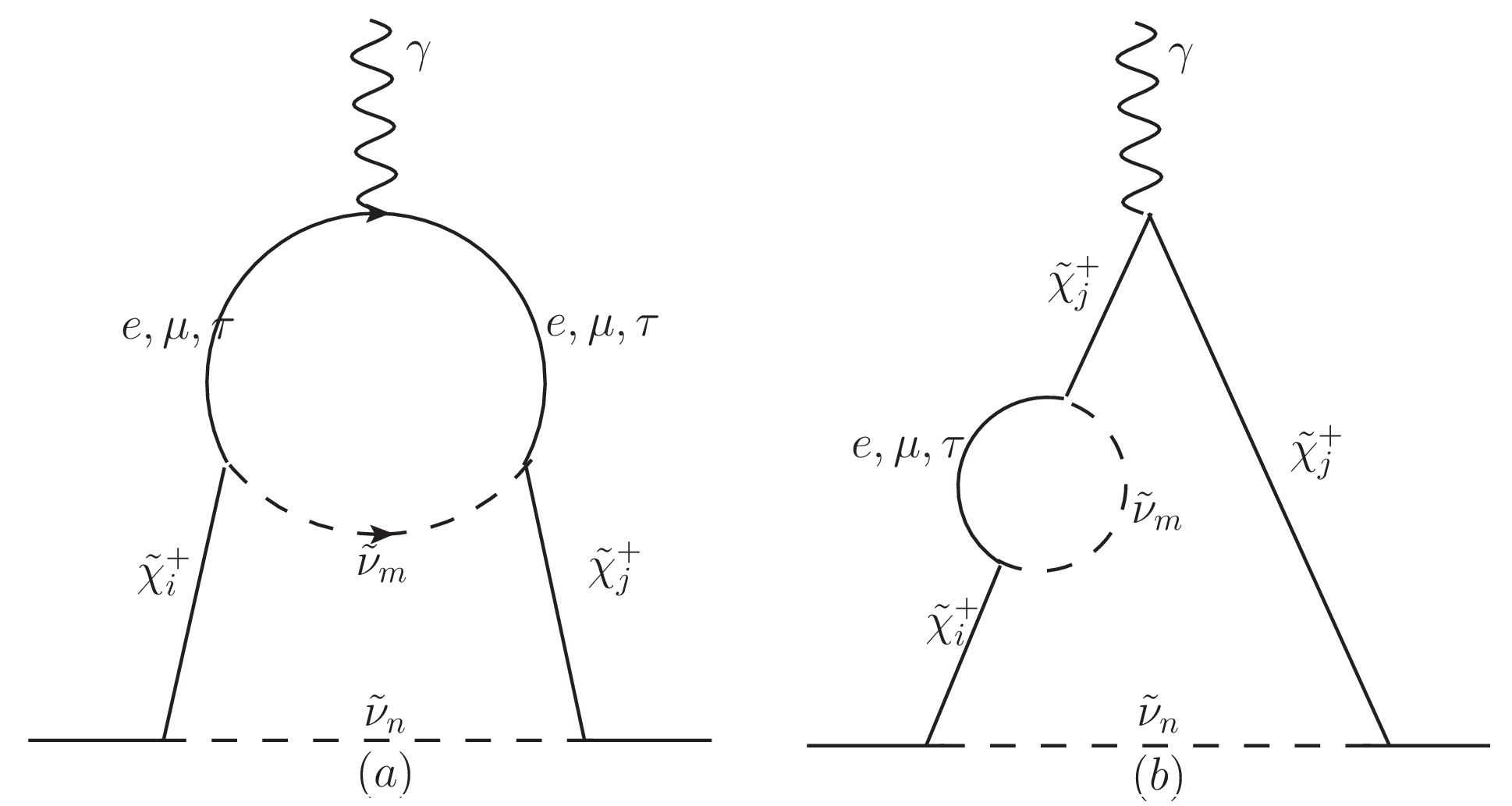}
    \caption{Supersymmetric two-loop Barr-Zee diagrams counterparts contributing to electron and muon EDM.}
    \label{2loop-barrzee-SUSY}
\end{figure}

Additionally, since the neutrino Yukawa couplings are sources of CP-violation because they are complex, supersymmetry makes sneutrino interactions complex as well, so they have contributions to fermions EDM via the diagrams shown in figure \ref{2loop-barrzee-SUSY}. In general, sneutrino can change chargino flavor although such processes have not  taken into account since only the main contribution is considered.

\begin{figure}[H]
    \centering
    \includegraphics[scale=0.55]{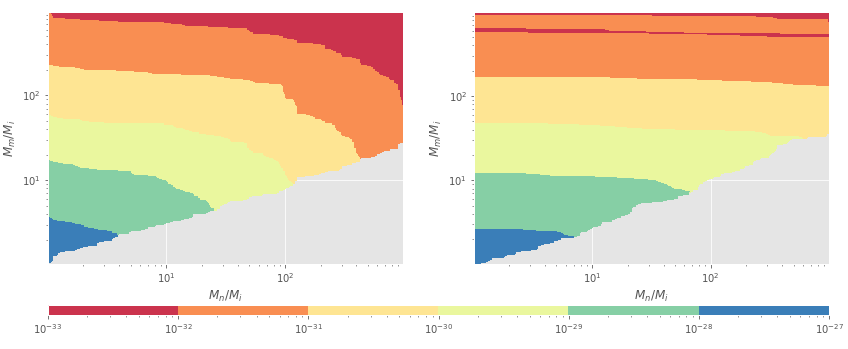}
    \caption{Contribution to electron and muon EDM for $M_{i}= 1$ TeV as a function of the sneutrino masses.}
    \label{newtwoloop}
\end{figure}

This kind of diagrams have been previously studied in \cite{yamanaka4} to the first order of the gauge boson momentum, and their importance due to potential large contributions to EDM have been discussed previously in \cite{refyamanaka4}. The general expression for the EDM contribution can be found in the appendices, which can also be used for degenerate chargino masses. In our case, since there is a light particle in the inner loop, their masses can be neglected in comparison to chargino and sneutrino masses. The contributions are shown in figure \ref{newtwoloop} and the expressions for the EDM, in the $m_{f},m_{e},m_{\mu} \ll M_{i}$ limit,  for diagram in  figure \ref{2loop-barrzee-SUSY}a ($d_{v}^{in \; jm}$) and for diagram in  figure \ref{2loop-barrzee-SUSY}b ($d_{p}^{in \; jm}$) for $i=j$ are given by:

\begin{align}
    -d_{v}^{in\; im}&=\frac{2}{(4\pi)^{4}}  M_{i}^{3} Im[g_{s}^{in} g_{p}^{in \;*}] (|Y_{p}^{im}|^{2}-|Y_{s}^{im}|^{2}) \int_{0}^{1} dx \int_{0}^{1} d\alpha  x^{2}h_{0}(x,\alpha) \nonumber\\
    &-\frac{1}{(4\pi)^{4}} M_{i} Im[g_{s}^{in}g_{p}^{*\; in}](|Y_{p}^{im}|^{2}-|Y_{s}^{im}|^{2}) \int_{0}^{1}dx\int_{0}^{1}d\alpha  \Bigg[ 2(3\alpha x+x-2) f_{0}(x,\alpha)  \nonumber\\
    &  +\frac{ x(2-x) m_{n}^2}{(1-x)M_{i}^2 - m_{n}^2}\Bigg[ (3\alpha-2) f(x,\alpha)\Big\rvert_{m_{o}=m_{f}=0}  + x(1-\alpha ) M_{i}^{2}g_{0}(x,\alpha) \Bigg] \Bigg] 
\end{align}

\footnotesize
\begin{align}
    -d_{p}^{in \; im}&=\frac{12}{(4\pi)^{4}} \int_{0}^{1} d\alpha \; \rho(\alpha) \int_{0}^{1} dx \int_{0}^{1-x} dy \; \Big[Im[g_{s}^{in} g_{p}^{in\;*}] (|Y_{p}^{im}|^{2}- |Y_{s}^{im}|^{2}) M_{i} (\alpha - 2) (1-x+y) \Big]  \tilde{f}_{0}(x,y) \nonumber\\
    & +\frac{2}{(4\pi)^{4}} \int_{0}^{1} d\alpha \; \rho(\alpha) \int_{0}^{1} dx \int_{0}^{1-x} dy \;  \Big[Im[g_{s}^{in} g_{p}^{in \; *}] (|Y_{p}^{im}|^{2} - |Y_{s}^{im}|^{2}) M_{i}^3 (\alpha - 2)(3y - 2)\Big]  \tilde{g}_{0}(x,y) \nonumber\\
    &+ \frac{2}{(4\pi)^{4}} \int_{0}^{1} d\alpha \; \rho(\alpha) \int_{0}^{1} dx \int_{0}^{1-x} dy \; \Big[Im[g_{s}^{in} g_{p}^{in\;*}](|Y_{p}^{im}|^{2}-|Y_{s}^{im}|^{2}) M_{i}^5(\alpha-2)(x+y) \Big]  \tilde{h}_{0}(x,y)
\end{align}

\normalsize
\noindent
where $g_{s(p)}^{in}$ is the scalar (pseudoscalar) coupling of the external particle with mass $m_{o}=m_{e}, m_{\mu}$ to the $i$-th chargino and the $n$-th sneutrino running into the outer loop, $Y_{s(p)}^{im}$ is the scalar (pseudoscalar) coupling of the $i$-th chargino to the inner fermion of mass $m_{f}$, where $f=e,\mu,\tau$, and the $m$-th sneutrino running into  the inner loop,  $i,j=1,2,3.$ label the chargino eigenstates of mass $M_{i}$, $m (n)=1,...9$ labels the sneutrino mass eigenstate running in the internal and external loop respectively with mass $m_{m(n)}$, However, we are not considering chargino flavor changes so $i=j$. Additionally, $h_{0}(x,\alpha)$, $f_{0}(x,\alpha)$, $g_{0}(x,\alpha)$, $\tilde{f}_{0}(x,\alpha)$, $\tilde{g}_{0}(x,\alpha)$ and $\tilde{h}_{0}(x,\alpha)$ are the functions shown in the appendices, $h(x,\alpha)$ $f(x,\alpha)$, $g(x,\alpha)$, $\tilde{f}(x,\alpha)$, $\tilde{g}(x,\alpha)$ and $\tilde{h}(x,\alpha)$ evaluated at $m_{o}=m_{f}=0$ respectively.

In general, it was found that the EDM contributions developed a singularity in the internal (external) loop located at $x=1-\frac{m_{m(n)}}{M_{i}} + \mathcal{O}\left(\frac{m_{f(o)}}{M_{i}}\right)$ that makes the integral divergent, implying that sneutrinos must be heavier than charginos. Likewise, it was found that $m_{m} \neq m_{n}$ and the diagram in figure \ref{newtwoloop}b was divergent when $m_{m}^{2}<M_{i}m_{n}$, reason why only a part of the graph is shown, and and resulted in the requirement of no denegerate sneutrino masses. Moreover, since the electron and muon masses were negligible in the calculations, such contributions are valid for both particles. Finally, the EDM can be rewritten as proportional to  $M_{i}^{-1}$ times a function depending on the ratios $\frac{m_{m}}{M_{i}}$ and $\frac{m_{n}}{M_{i}}$. In figure \ref{newtwoloop} is shown for the particular case when $M_{i}=1\; TeV$ which implies the possibility of EDM above the experimental upper bound (blue and dark green zone). Thus, it gives a  chargino mass lower bound of $M_{i}>10^{5}$ GeV so all possible values lie under the experimental upper bound.

\section{Conclusions}
Although the non-universal extension to the MSSM has proven to be compatible with SM phenomenology, the electron and muon EDM was studied by considering additional CP violating sources coming from exotic neutrinos as well as their supersymmetric counterpart, being the mass of each particle a free parameter in the model. Complex Yukawa couplings arise to match the CP violating phase of the PMNS matrix, which makes sneutrinos to have complex couplings as well as exotic neutrinos mass eigenstates due to the inverse seesaw mechanism rotation. From the one-loop contributions one can see that if any particle into the loop has mass in the TeV scale the other must be heavier by at least two orders of magnitude. Besides, electron and muon EDM upper bound implies that charged scalars and exotic neutrinos cannot have similar masses. From the two-loop contributions, we have seen that they are comparable with the one-loop values, the convergence of the EDM form factor integral forbids degenerate sneutrino masses and it requires sneutrinos to be heavier than charginos. However, despite the model is able to predict a small EDM for the electron and muon, the current experimental upper bound provide important restrictions on chargino and sneutrino masses.
\appendix
\section{Calculation of diagram \ref{2loop-barrzee-SUSY}a}

The amplitude of diagram shown in figure \ref{2loop-barrzee-SUSY}a defines the vertex function $\bar{u}(p)  \Gamma_{v\; \mu}^{in\;jm}u(p')$ which can be written as:
\small
\begin{align}
    \Gamma_{v\; \mu}^{in\;jm} &= -\int \frac{d^{4}k}{(2\pi)^{4}}  \frac{1}{(p-k)^{2}-m^{2}}(g_{s}^{in}+i g_{p}^{in}\gamma^{5})\frac{\slashed{k}+M}{k^{2}-M_{i}^{2}}\mathcal{M}_{int} \frac{\slashed{k}-\slashed{q}+M_{j}}{(k-q)^{2}-M_{j}^{2}}(g_{s}^{* \; jm}-i g_{p}^{* \; jm}\gamma^{5})
\end{align}

\normalsize
\noindent
where the internal loop amplitude is given in Eq. \ref{Mint} and has a logarithmic superficial degree of divergence. The internal loop momentum $l$ is decoupled by the shift $r=l-xk -zq$ being $x$ and $z$ Feynman parameters such that $x+y+z=1$, leading to:

\small
\begin{align}
    \mathcal{M}_{int}&=(Y_{s}^{in}+ i Y_{p}^{in} \gamma^{5})\int \frac{d^{4}l}{(2\pi)^{4}}\frac{1}{(k-l)^{2}-m^{2}}\frac{\slashed{l}+m_{f}}{l^{2}-m_{f}^{2}}\gamma_{\mu}   \frac{\slashed{l}-\slashed{q}+m_{f}}{(l-q)^{2}-m_{f}^{2}} (Y_{s}^{*\; jm}-i Y_{p}^{* \; jm} \gamma^{5}) \label{Mint}\\
    &=\int dx dy dz \delta(x+y+z-1) \int \frac{d^{4}r}{(2\pi)^{4}} (Y_{s}^{in}+ i Y_{p}^{in} \gamma^{5})\Bigg[\frac{\slashed{r}\gamma_{\mu}\slashed{r}}{(r^{2}-\Delta_{I})^{3}} \nonumber \\ 
    &\;\;\;\;\;\;\;\;+ \frac{(x\slashed{k}+z\slashed{q})\gamma_{\mu}(x\slashed{k}+(z-1)\slashed{q})+m_{f}^{2}\gamma_{\mu}}{(r^{2}-\Delta_{I})^{3}}  +\frac{2m_{f}(xk_{\mu}+zq_{\mu})-m_{f}\gamma_{\mu}\slashed{q}}{(r^{2}-\Delta_{I})^{3}} \Bigg](Y_{s}^{im \; *}-i Y_{p}^{im \; *} \gamma^{5})
\end{align}

\normalsize
\noindent
\begin{figure}
    \centering
    \includegraphics[scale=0.5]{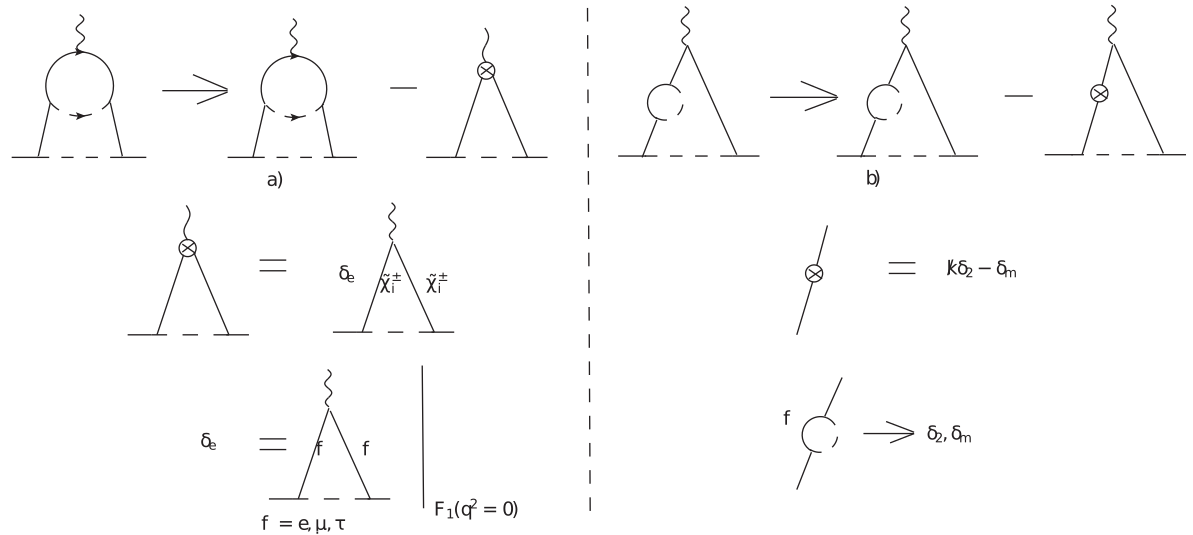}
    \caption{Counterterm diagrams necessary for the UV convergence of EDM contributions from diagrams in figure \ref{2loop-barrzee-SUSY}.}
    \label{corrections}
\end{figure}

\noindent
being $\Delta_{I}=-x(1-x)k^{2}-z(1-z)q^{2}+2xz k \cdot q +xm^{2}+(1-x)m_{f}^{2}$. The first term diverges while the second and third one converges. To remove the divergence in the first term, we need to consider an additional counterterm diagram as shown in figure \ref{corrections}a where $\delta_{e}=-F_{1}(q^{2}=0)$ is the electric charge form factor coming from the subdiagram involving light fermions. Considering only the divergent part, we arrive to:

\small
\begin{align}
    \mathcal{M}_{int}^{div}&=\int dx dy dz \delta(x+y+z-1) \int \frac{d^{4}r}{(2\pi)^{4}} (Y_{s}^{in}+ i Y_{p}^{in} \gamma^{5}) \Bigg[\frac{\slashed{r}\gamma_{\mu}\slashed{r}}{(r^{2}-\Delta_{I})^{3}} \Bigg](Y_{s}^{im \; *}-i Y_{p}^{im \; *} \gamma^{5}) + \delta_{e} \nonumber\\
    &=\int dx dy dz \delta(x+y+z-1) \frac{i \gamma_{\mu}}{2(4\pi)^{2}}\Bigg[ (|Y_{p}^{im}|^{2}-|Y_{s}^{im}|^{2}) \log\left(\frac{\Delta_{I}}{\Delta_{c}}\right) \nonumber \\
    & \;\;\;\;\;\;\;\;\;\;\;\;\;\;\;\;\;\;\;\;\;\;\;\;\;\;\;\;\;\;\;\;\;\;\;\;\;\;\;\;\;\;\;\;\;\;\;\;\;\;\;\;\;\;\;\;\;\;\;\;\;\;\;- \frac{|Y_{s}^{im}|^{2}(M_{i}x+m_{f})^{2}-|Y_{p}^{im}|^{2}(M_{i}x-m_{f})^{2}}{\Delta_{c}}\Bigg]
\end{align}
 
 \normalsize
 \noindent
 where $\Delta_{c}=-x(1-x)M^{2} +xm^{2}+(1-x)m_{f}^{2}$. We can see that the second term is $k$ independent so it factorizes when integrating the outer loop. Besides, to simplify the expressions we apply the condition $q^{2}=0$ to extract the EDM contribution before integrating over Feynmann parameters. Nevertheless, the first term is developed by doing the $y$-integration to remove the Dirac delta function, then we integrate by parts on the $z$ variable and then again on the $x$ variable, resulting in:

\begin{align}
    \mathcal{M}_{int}^{div}&=-\frac{i\gamma_{\mu}}{4(4\pi)^{2}} (|Y_{p}^{im}|^{2}-|Y_{s}^{im}|^{2}) \int_{0}^{1} dx \frac{x(2-x)^{2} (m_{n}^2 x^2 - (1-x)^2 m_{f}^2 )}{( -x(1-x)M_{i}^2 + x m_{n}^2 + (1-x)m_{f}^2)}\times \nonumber\\
    &\;\;\;\;\;\;\;\;\;\;\;\;\;\;\;\;\;\;\;\;\;\;\;\;\;\;\;\;\;\;\;\;\;\;\;\;\;\;\;\;\;\;\;\times\frac{(k^2 - 2 k\cdot q - M_{i}^2)}{(-x (1-x)k^2 + 2x(1-x) k\cdot q  + xm_{n}^2 + (1-x)m_{f}^2)}\nonumber\\
    &-\frac{i\gamma_{\mu}}{2(4\pi)^{2}} (|Y_{p}^{im}|^{2}-|Y_{s}^{im}|^{2}) \int_{0}^{1} dx \int_{0}^{1-x} dz \; \frac{2xz k \cdot q}{-x(1-x)k^{2}+2xz k \cdot q +xm_{n}^{2}+(1-x)m_{f}^{2}} \nonumber \\
    & -\frac{i\gamma_{\mu}}{2(4\pi)^{2}} \int_{0}^{1} dx (1-x)  \frac{|Y_{s}^{im}|^{2}(M_{i}x+m_{f})^{2}-|Y_{p}^{im}|^{2}(M_{i}x-m_{f})^{2}}{-x(1-x)M_{i}^{2} +xm_{n}^{2}+(1-x)m_{f}^{2}} .
\end{align}

Then, the integration over the $k$ momentum is straightforward and the  EDM contribution is extracted by using the projector given in \cite{EDMprojector} which is already implemented in Package-X \cite{package-x} and taking only CP non-invariant terms, giving as a final result:

\small
\begin{align}
    -d_{v}^{in\; im}&=\frac{1}{(4\pi)^{4}} M_{i} Im[g_{s}^{in}g_{p}^{in \;*}](|Y_{p}^{im}|^{2}-|Y_{s}^{im}|^{2}) \int_{0}^{1} dx \int_{0}^{1} d\alpha \frac{x(1-x)(2-x) \left(\frac{m_{f}^2}{x^{2}} -\frac{m_{n}^2}{(1-x)^{2}} \right)}{M_{i}^2 - \frac{m_{n}^2}{1-x} - \frac{m_{f}^2}{x}}\times\nonumber\\
    & \;\;\;\;\;\;\;\;\;\;\;\;\;\;\;\;\;\;\;\;\;\;\;\;\;\;\;\;\;\;\;\;\;\;\;\;\;\;\;\;\;\;\;\;\;\;\;\;\;\;\;\;\; \times \Bigg[ \frac{3\alpha-2}{1-x} f(x,\alpha)  + x(1-\alpha )(M_{i}^{2} - \alpha^{2} m_{o}^{2})g(x,\alpha) \Bigg] \nonumber\\ 
    & +\frac{2}{(4\pi)^{4}}  M_{i} Im[g_{s}^{in} g_{p}^{in \;*}] \int_{0}^{1} dx \int_{0}^{1} d\alpha  \Big[ (|Y_{p}^{im}|^{2}-|Y_{s}^{im}|^{2})(M_{i}^{2}x^{2}+m_{f}^{2})\nonumber\\
    &\;\;\;\;\;\;\;\;\;\;\;\;\;\;\;\;\;\;\;\;\;\;\;\;\;\;\;\;\;\;\;\;\;\;\;\;\;\;\;\;\;\;\;\;\;\;\;\;\;\;\;\;\;\;\;\;\;\;\;\;\;\;\;\;\;\;\;\;\;\;\;\;\;\;\;\;\;-(|Y_{s}^{im}|^{2}+|Y_{p}^{im}|^{2})(2M_{i}m_{f}x)\Big]h(x,\alpha) \nonumber\\
    &-\frac{2}{(4\pi)^{4}} M_{i} Im[g_{s}^{in}g_{p}^{*\; in}](|Y_{p}^{im}|^{2}-|Y_{s}^{im}|^{2}) \int_{0}^{1}dx\int_{0}^{1}d\alpha  (3\alpha x+x-2) f(x,\alpha)  \nonumber \\
     &-\frac{2}{(4\pi)^{4}}m_{f} \Big((|g_{p}^{in}|^{2}+|g_{s}^{in}|^{2})Im[Y_{s}^{im}Y_{p}^{*\; im}]-Im[g_{s}^{in}g_{p}^{*\; in}](|Y_{p}^{im}|^{2}+|Y_{s}^{im}|^{2})\Big)\nonumber\\
    &\;\;\;\;\;\;\;\;\;\;\;\;\;\;\;\;\;\;\;\;\;\;\;\;\;\;\;\;\;\;\;\;\;\;\;\;\;\;\;\;\;\;\;\;\;\;\;\;\;\;\;\;\;\;\;\;\;\;\;\;\;\;\;\;\;\;\;\;\;\;\;\;\;\;\;\;\;\;\;\;\;\;\;\;\;\;\;\;\;\;\;\;\;  \int_{0}^{1}dx\int_{0}^{1}d\alpha (3\alpha -1) f(x,\alpha) \nonumber\\
     &+\frac{1}{(4\pi)^{4}}\int_{0}^{1}dx \int_{0}^{1} d\alpha \Bigg[ M_{i} m_{o}^{2}Im[g_{s}^{in}g_{p}^{*\; in}] (|Y_{p}^{im}|^{2}-|Y_{s}^{im}|^{2}) x^{2} \alpha^{2}(\alpha -1) \nonumber \\
     &\;\;\;\;\;\;\;\;\;\;\;\;\;\;\;\;\;\; +  m_{f}(|g_{p}^{in}|^{2} + |g_{s}^{in}|^{2})Im[Y_{s}^{im} Y_{p}^{*\; im}] (\alpha^2 m_{o}^2 ((\alpha - 2) x + 1) + M_{i}^2 (1 - \alpha x)) \nonumber \\
     &\;\;\;\;\;\;\;\;\;\;\;\;\;\;\;\;\;\;+ 2 m_{f} M_{i} m_{o}(|g_{p}^{in}|^{2} - |g_{s}^{in}|^{2})Im[Y_{s}^{im} Y_{p}^{*\; im}] \alpha( x + 1) \nonumber \\
     &\;\;\;\;\;\;\;\;\;\;\;\;\;\;\;\;\;\; + m_{f} Im[g_{s}^{in} g_{p}^{*\; in}](|Y_{p}^{im}|^2 + |Y_{s}^{im}|^2) (M_{i}^2 (1 - \alpha x) - \alpha^2 m_{o}^2 ((\alpha - 2) x + 1)) \Bigg] g(x,\alpha) \nonumber
\end{align}

\noindent
where $m_{o}$ is the external particle mass. Besides, the first three line comes from the divergent term, the fourth and fifth line , proportional to $f(x,\alpha)$, from the terms proportional to $t^{2}$, being $t$ the shifted momentum that  allows to decouple the $k$ integration, and the last lines come from terms proportional to $t^{0}$. Finally, the functions $f$, $g$ and $h$ are given by:

\small
\begin{align}
f(x,\alpha)&= \frac{(\alpha-1)}{M_{i}^2-\frac{m_{f}^2}{x}-\frac{m_{m}^2}{1-x} }\left[ 1 +\frac{ \frac{m_{m}^{2}}{1-x}+\frac{m_{f}^2}{x}-\alpha m_{o}^{2}-\frac{\alpha}{\alpha -1} m_{n}^{2}}{M_{i}^2-\frac{m_{f}^2}{x}-\frac{m_{m}^2}{1-x}} \log\left(\frac{\frac{m_{f}^2}{x}+\frac{m_{m}^{2}}{1-x}-\alpha  m_{o}^{2} +\frac{\alpha}{1-\alpha} m_{n}^{2} }{M_{i}^2-\alpha m_{o}^2 +\frac{\alpha}{1-\alpha } m_{n}^2}\right)  \right] \nonumber\\
g(x,\alpha)&= \frac{x^{-1}}{M_{i}^2 -\frac{m_{f}^2}{x}-\frac{m_{m}^2}{1-x}} \left[\frac{1}{M_{i}^2 - \alpha m_{o}^2+\frac{\alpha}{1-\alpha} m_{n}^2} \right. \nonumber \\
&\;\;\;\;\;\;\;\;\;\;\;\;\;\;\;\;\;\;\;\;\;\;\;\;\;\;\;\;\;\;\;\;\;\;\;\;\;\;\;\;\;\;\;\;\;\;\;\;\;\;\;\;\;\;\left. +\frac{1}{M_{i}^2-\frac{m_{f}^2}{x} -\frac{m_{m}^2}{1-x} }\log  \left(\frac{\frac{m_{f}^2}{x} + \frac{m_{m}^2}{1-x}-\alpha  m_{o}^2 + \frac{\alpha}{1-\alpha} m_{n}^2}{M_{i}^2 -\alpha m_{o}^2 + \frac{\alpha}{1-\alpha} m_{n}^2}\right)\right] \nonumber\\
h(x,\alpha)&=\frac{x^{-1}\alpha^{-1}(1-\alpha)}{\left(M_{i}^{2} -\frac{m_{m}^{2}}{1-x}-\frac{m_{f}^{2}}{x}\right)\left(m_{o}^{2}-\frac{ m_{n}^{2}}{1-\alpha}-\frac{M_{i}^{2}}{\alpha}\right)} \nonumber
\end{align}
\normalsize

\section{Calculation of diagram \ref{2loop-barrzee-SUSY}b}

The amplitude of diagram shown in figure \ref{2loop-barrzee-SUSY}b defines the vertex function $\bar{u}(p)  \Gamma_{p\; \mu}^{in\;im}u(p')$ which can be written as:

\begin{align}
   \Gamma_{p\; \mu}^{in\;im} &=-\int \frac{d^{4}k}{(2\pi)^{4}}\frac{(g_{s}^{in}+ig_{p}^{in}\gamma^{5})}{(p-k)^{2}-m_{m}^{2}}\frac{\slashed{k}+M_{i}}{k^{2}-M_{i}^{2}}\mathcal{M}_{int} \frac{\slashed{k}+M_{i}}{k^{2}-M_{i}^{2}}\gamma_{\mu}\frac{\slashed{k}-\slashed{q}+M_{i}}{(k-q)^{2}-M_{i}^{2}}(g_{s}^{in \; *} -ig_{p}^{in \; *}\gamma^{5})
\end{align}

\noindent
where the internal loop is given by:

\begin{align}
    \mathcal{M}_{int}&=\int \frac{d^{4}l}{(2\pi)^{4}}(Y_{s}^{im}+iY_{p}^{im}\gamma^{5})\frac{\slashed{l}+m_{f}}{(l^{2}-m_{f}^{2})((k-l)^{2}-m_{n}^{2})}(Y_{s}^{im\; *}-iY_{p}^{im\; *}\gamma^{5}).
\end{align}

The internal loop integral diverges so we need to consider the counterterm diagram shown in Figure \ref{corrections}b whose amplitude is given by:

\begin{align}
    \mathcal{M}_{int}^{c}&=(Y_{s}^{im}+iY_{p}^{im}\gamma^{5}) (\slashed{k}\delta_{2} - \delta_{m}) (Y_{s}^{im\; *}-iY_{p}^{im\; *}\gamma^{5})
\end{align}

\noindent
where $\delta_{2}$ and $\delta_{m}$ are one-loop renormalization constants related to chargino wave function and mass respectively, given by:

\small
\begin{align*}
    \delta_{2}&=\frac{i}{(4\pi)^{2}} \int_{0}^{1} d\alpha (1-\alpha) \left( \frac{2}{\epsilon}-\log(-\alpha(1-\alpha)M_{i}^{2}+\alpha m_{f}^{2}+(1-\alpha)m_{m}^{2}) -\gamma +  \log(4\pi) + \mathcal{O}(\epsilon)\right) \nonumber \\
    \delta_{m}&=\frac{-i}{(4\pi)^{2}} \int_{0}^{1} d\alpha \; m_{f} \left( \frac{2}{\epsilon}-\log(-\alpha(1-\alpha)M_{i}^{2}+\alpha m_{f}^{2}+(1-\alpha)m_{m}^{2}) -\gamma +  \log(4\pi) + \mathcal{O}(\epsilon)\right) \nonumber \\
\end{align*}

\normalsize
\noindent
where $\gamma$ is the Euler-Mascheroni constant. Then, the renormalized internal loop is given by:

\begin{align}
    \mathcal{M}_{int}^{R}&=\mathcal{M}_{int}-\mathcal{M}_{int}^{c} \nonumber \\
    &=\frac{i}{(4\pi)^{2}} \int_{0}^{1} d\alpha \; ((1-\alpha)\slashed{k}+m_{f}) \log \left( \frac{-\alpha(1-\alpha)M_{i}^{2}+\alpha m_{f}^{2}+(1-\alpha)m_{m}^{2}}{-\alpha(1-\alpha)k^{2}+\alpha m_{f}^{2}+(1-\alpha)m_{m}^{2}} \right) \nonumber\\
    &= \frac{i}{2(4\pi)^{2}} \int_{0}^{1} d\alpha \frac{m_{f}^{2}-m_{m}^{2}(1-\alpha)^{2}}{(1-\alpha)(-\alpha(1-\alpha)M_{i}^{2}+\alpha m_{f}^{2}+(1-\alpha)m_{m}^{2})} \frac{(k^{2}-M_{i}^{2})((2-\alpha)\slashed{k}+2m_{f})}{k^{2}-\frac{m_{f}^{2}}{1-\alpha}-\frac{m_{m}^{2}}{\alpha}} \nonumber \\
    & \equiv \frac{i}{2(4\pi)^{2}} \int_{0}^{1} d\alpha \rho(\alpha) \frac{(k^{2}-M_{i}^{2})((2-\alpha)\slashed{k}+2m_{f})}{k^{2}-\frac{m_{f}^{2}}{1-\alpha}-\frac{m_{m}^{2}}{\alpha}}
\end{align}
\noindent
where the integration over the internal momentum was done by the shift $r=l-(1-x)k$ and in spite of the divergences on the $k$ momentum integral, such divergences do not contribute to the EDM form factor, so we can ignore electric charge renormalization up to two-loop level. After doing the $t=k-xp-yq$ shift and implementing the EDM form factor projector on Feyncalc, the final contribution can be written as:

\small
\begin{align*}
    -d_{p}^{in \; im}&=-\frac{12}{(4\pi)^{4}} \int_{0}^{1} d\alpha \; \rho(\alpha) \int_{0}^{1} dx \int_{0}^{1-x} dy \; \bigg[ Im[g_{s} g_{p}^{*}] \Big[(|Y_{p}|^{2} + |Y_{s}|^{2}) m_{f} (1 - 2y) + \nonumber \\
    & \;\;\;\;\;\;\;\;\;\;\;\;\;+(|Y_{p}|^{2}- |Y_{s}|^{2}) (\alpha - 2) M_{i} (x-y + 1) \Big] +  (|g_{p}|^{2} + |g_{s}|^{2}) Im[Y_{s}Y_{p}^{*}] m_{f} (2 y - 1) \bigg] \tilde{f}(x,y) \\
    & +\frac{2}{(4\pi)^{4}} \int_{0}^{1} d\alpha \; \rho(\alpha) \int_{0}^{1} dx \int_{0}^{1-x} dy \; \bigg[  Im[g_{s} g_{p}^{*}] \Big[(|Y_{p}|^{2}+ |Y_{s}|^{2}) 2 m_{f} (3 M_{i}^2 (x + y) +  mo^2 x^2 (3 - 4y)) +\nonumber \\
    & \;\;\;\;\;\;\;\;\;\;\;\;\;\;\;\;\;\;\;\;+ (|Y_{p}|^{2} - |Y_{s}|^{2}) (\alpha - 2) (M_{i}^3 (3y - 2) + 2 M_{i} mo^2 x^2 (2x - 2y + 3))\Big]+\nonumber \\
    & \;\;\;\;\;\;\;\;\;\;\;\;\;\;\;\;\;\;\;\;\;\;\;\;\;\;\;\;\;\; + 2 m_{f} (|g_{p}|^{2}+ |g_{s}|^{2}) Im[Y_{s} Y_{p}^{*}](M_{i}^2 (2-3 y) + mo^2 x^2 (4y - 3)) \bigg] \tilde{g}(x,y) \\
    &+ \frac{2}{(4\pi)^{4}} \int_{0}^{1} d\alpha \; \rho(\alpha) \int_{0}^{1} dx \int_{0}^{1-x} dy \; \bigg[ 2 m_{f}(M_{i}^2-m_{o}^2 x^2)^2 (1-y)(|g_{p}|^{2}+|g_{s}|^{2}) Im[Y_{s} Y_{p}^{*}] +  \nonumber \\
    & \;\;\;\;\;\;\;\;\;\;\;+(M_{i}^{2}-m_{o}^{2} x^{2}) Im[g_{s} g_{p}^{*}] \Big[ (|Y_{p}|^{2}+|Y_{s}|^{2}) 2 m_{f} (M_{i}^2 (2 x+y+1)-m_{o}^2 x^2 (y-1)) + \\
    & \;\;\;\;\;\;\;\;\;\;\;\;\;\;\;\;\;\;\;\;\;\; +(|Y_{p}|^{2}-|Y_{s}|^{2}) (\alpha-2) (M_{i}^3 (x+y)+M_{i} m_{o}^2 x^2 (x-y+2)) \Big] \bigg] \tilde{h}(x,y)
\end{align*}

\noindent
where the function $\rho$, $\tilde{f}$, $\tilde{g}$ and $\tilde{h}$ are given by:

\begin{align*}
    \rho(\alpha)&=\frac{m_{f}^{2}-m_{m}^{2}(1-\alpha)^{2}}{(1-\alpha)(-\alpha(1-\alpha)M_{i}^{2}+\alpha m_{f}^{2}+(1-\alpha)m_{m}^{2})} \\
    \tilde{f}(x,y)&= \frac{1-x-y}{\left(M_{i}^{2}-\frac{m_{f}^{2}}{1-\alpha}-\frac{m_{m}^{2}}{\alpha}\right)}+\frac{x(1-x)\left(m_{o}^2-\frac{m_{n}^2}{1-x} -\frac{y}{x(1-x)} M_{i}^2 -\frac{1-x-y}{x(1-x)}\left(\frac{m_{f}^{2}}{1-\alpha}+\frac{m_{m}^{2}}{\alpha}\right) \right)}{\left(M_{i}^{2}-\frac{m_{f}^{2}}{1-\alpha}-\frac{m_{m}^{2}}{\alpha}\right)^{2}} \times  \nonumber\\
    &\;\;\;\;\;\;\;\;\;\;\;\;\;\;\;\;\;\;\;\;\;\;\;\;\;\;\;\;\;\;\;\;\;\;\;\;\;\;\;\;\;\;\;\;\;\;\;\;\;\;\;\;\;\;\;\;\;\times\log \left( \frac{m_{o}^{2}-\frac{m_{n}^{2}}{1-x} - \frac{M_{i}^{2}}{x}}{m_{o}^{2}-\frac{m_{n}^{2}}{1-x}-\frac{y}{x(1-x)}M_{i}^{2} - \frac{1-x-y}{x(1-x)}\left(\frac{m_{f}^{2}}{1-\alpha}+\frac{m_{m}^{2}}{\alpha}\right)} \right) \\
    \tilde{g}(x,y)&=\frac{1}{\left(M_{i}^{2}-\frac{m_{f}^{2}}{1-\alpha}-\frac{m_{m}^{2}}{\alpha}\right)^{2}} \log \left( \frac{m_{o}^{2}-\frac{m_{n}^{2}}{1-x} - \frac{M_{i}^{2}}{x}}{m_{o}^{2}-\frac{m_{n}^{2}}{1-x}-\frac{y}{x(1-x)}M_{i}^{2} - \frac{1-x-y}{x(1-x)}\left(\frac{m_{f}^{2}}{1-\alpha}+\frac{m_{m}^{2}}{\alpha}\right)} \right) \nonumber\\
    &\;\;\;\;\;\;\;\;\;\;\;\;\;\;\;\;\;\;\;\;\;\;\;\;\;\;\;\;\;\;\;\;\;\;\;\;\;\;\;\;\;\;\;\;\;\;\;\;\;\;+ \frac{1-x-y}{x(1-x)\left(M_{i}^{2}-\frac{m_{f}^{2}}{1-\alpha}-\frac{m_{m}^{2}}{\alpha}\right)\left(m_{o}^{2}- \frac{m_{n}^{2}}{1-x}-\frac{M_{i}^{2}}{x}\right)}\\
    \tilde{h}(x,y)&=\frac{(1-x-y)^{2}}{ 2 (-x(1-x))^{3}\left( m_{o}^2 -\frac{m_{n}^2}{1-x} -\frac{M_{i}^2}{x}\right)^2 \left(m_{o}^2 -\frac{m_{n}^2}{1-x}  - \frac{y}{x(1-x)} M_{i}^2 -\frac{1-x-y}{x(1-x)} \left(\frac{m_{f}^{2}}{1-\alpha}+\frac{m_{m}^{2}}{\alpha}\right) \right)}
\end{align*}

\end{document}